\documentclass{article}
\pdfoutput=1

\PassOptionsToPackage{numbers, compress}{natbib}  

\usepackage[preprint]{neurips_2026}

\usepackage[utf8]{inputenc} % allow utf-8 input
\usepackage[T1]{fontenc}    % use 8-bit T1 fonts
\usepackage{hyperref}       % hyperlinks
\usepackage{url}            % simple URL typesetting
\usepackage{booktabs}       % professional-quality tables
\usepackage{amsfonts}       % blackboard math symbols
\usepackage{nicefrac}       % compact symbols for 1/2, etc.
\usepackage{microtype}      % microtypography
\usepackage{xcolor}         % colors 
\usepackage{amsmath}
\usepackage{graphicx}   % for \resizebox   
\usepackage{pifont}
\usepackage{MnSymbol}
\usepackage{subfigure}

\usepackage{caption} 
\usepackage{changes}
\usepackage{comment}
\usepackage{enumitem}

\newcommand{\AeRot}{AeRot}            % kernel name -- change here only
\newcommand{\Ntrain}{N_{\rm train}}

\newcommand{\matern}{Mat\'{e}rn}
\newcommand{\Nwindow}{N_{\rm window}}
\newcommand{\nahead}{n_{\rm ahead}}
\newcommand{\TL}{T_L}
\newcommand{\Rini}{R_{\rm ini}}          % principled initial guess for R

\title{Structured Quantum Kernels for Chaotic Forecasting}

\author{%
  Zhihui Wang$^{1,3}$ \quad Sujit Roy$^{1,2}$ \quad Ata Akbari$^{3}$ \quad
  Manil Maskey$^{1}$ \quad Rahul Ramachandran$^{1}$\\[6pt]
  $^1$IMPACT AI, Office of Data Science and Informatics (ODSI)/NASA MSFC, Huntsville, AL, USA\\
  $^2$The University of Alabama in Huntsville, Huntsville, AL, USA\\
  $^3$Research Institute for Advanced Computer Science (RIACS), Universities Space Research Association (USRA), USA\\[4pt]
  \texttt{zhihui.wang@nasa.gov} \qquad \texttt{sujit.roy@nasa.gov}
}

\begin{document}

\maketitle

\begin{abstract}

Quantum kernels promise exponentially large feature spaces, but expressive circuits render their Gram matrices uninformative, bandwidth tuning collapses them toward classical RBF, and the practical consensus is that quantum kernels add nothing on classical data.
We provide an answer to another productive question, an architectural one: whether the \emph{structure} of an encoding circuit can carry an inductive bias that tuned classical kernels lack.
We introduce AeRot, a quantum kernel that fuses amplitude encoding of $\ell^2$-normalized delay windows with a grouped single-qubit rotation layer assigning contiguous temporal blocks to each qubit, making the circuit delay-window-aware.
On Lorenz-63 non-autoregressive kernel ridge regression (KRR) with cross-validated bandwidths for 100 seeds, AeRot outperforms tuned RBF and Mat\'ern-5/2, with the advantage emerging at physical horizons $\geq 0.15$\,tu and reaching $+0.137$ mean $R^2$ at $0.25$\,tu (5 qubits, $N{=}32$).
Linear-stability analysis of the window tail localises the advantage to the locally unstable saddle-approach regime: AeRot wins 83\% of windows in the most unstable decile, with the quantum-win rate scaling monotonically with tail instability and a sign flip at the local stability boundary ($\bar\lambda_{\rm tail} \approx 0$).
The difficulty is fold-branch ambiguity: trajectories approaching the saddle are locally diverging, and Euclidean kernels struggle to resolve which lobe the trajectory will commit to.
Structural diagnostics confirm Gram matrices structurally distinct from tuned RBF (normalized Frobenius distance $\approx 0.44$) and strictly higher target-kernel alignment at every horizon ($95$--$100\%$ of seeds).
The gain is architectural: a finite-sample inductive-bias effect of temporally structured encoding on a folded attractor, with no computational-separation claim attached.
To our knowledge, this is the first mechanistic localisation of quantum-kernel advantage to a specific dynamical regime of a classical system.
\end{abstract}

\section{Introduction}
\label{sec:intro}
\begin{figure}[t]
  \centering
  \begin{minipage}[c]{0.5\linewidth}
    \centering
    {\bfseries a}\\[2pt]
    \includegraphics[width=\linewidth]{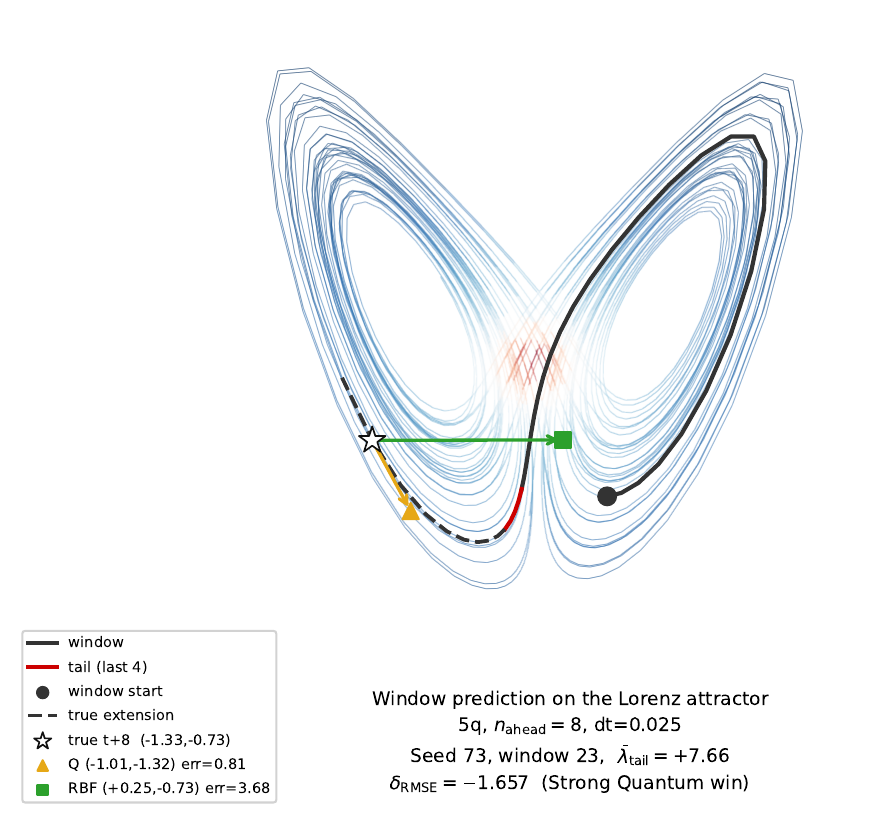}
  \end{minipage}%
  \hfill
  \begin{minipage}[c]{0.5\linewidth}
    \centering
    {\bfseries b}\\[1pt]
    \includegraphics[width=\linewidth]{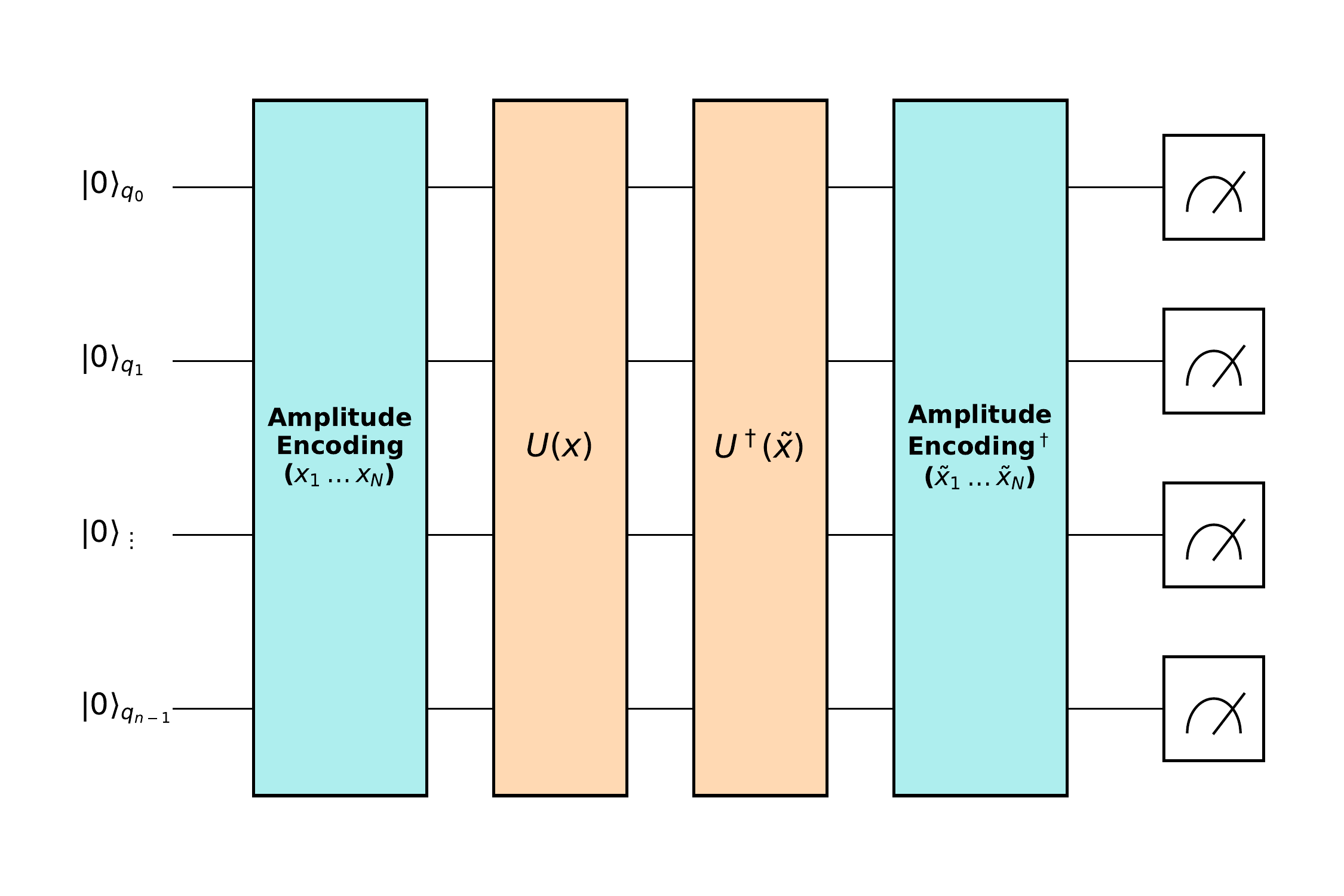}\\[-2pt]
    {\bfseries c}\\[1pt]
    \includegraphics[width=\linewidth]{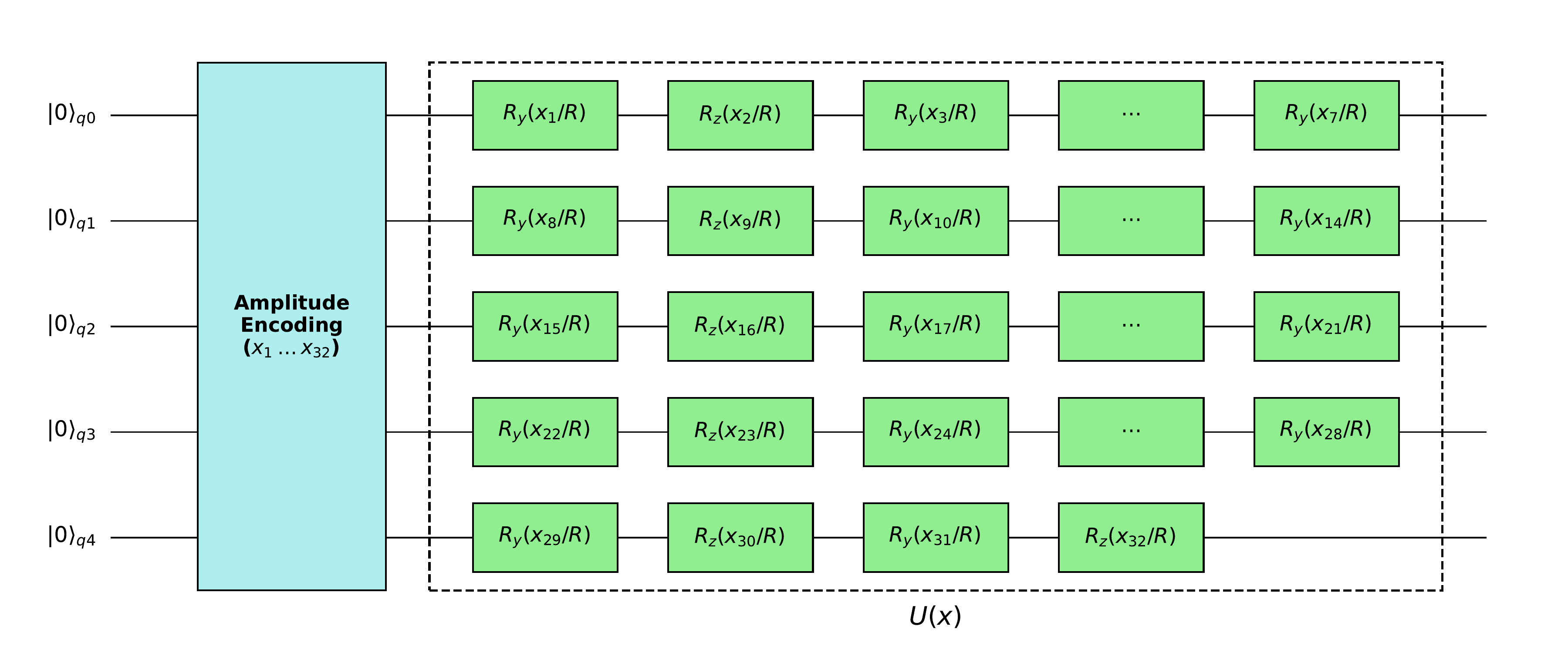}
  \end{minipage}
  \caption{\textbf{(a)}~A test window from the saddle-approach band
    of the Lorenz-63 attractor (seed 73, window 23).  Blue lines trace
    the full trajectory on the $xz$-plane; the black track is the
    32-step input window ending near the origin saddle. The $\star$ marks the true tangent at horizon $0.20$\,tu. The red piece is the window tail
    whose instability drives the hardness of the task.
    RBF (\textcolor{green!70!black}{$\blacksquare$},
    error $= 3.68$) is pulled toward the wrong branch of the fold;
    \AeRot{} (\textcolor{orange}{$\blacktriangle$}, error $= 0.81$) remains near the true
    target.
    \textbf{(b)}~The overlap circuit computing $\kappa(x,y)$ for
    general data $(x_1,\ldots,x_N)$ encoded in $n$ qubits.
    \textbf{(c)}~State preparation $U(x)$ on $n=5$ qubits.  Each
    qubit receives a contiguous temporal block of the window
    (colour-coded) via alternating $R_y/R_z$ gates; no inter-qubit
    entanglement beyond amplitude encoding.}
  \label{fig:overview}
  \vspace{-1em}
\end{figure}

Quantum kernels embed classical inputs into $2^n$-dimensional Hilbert space and
estimate state overlaps as similarity measures~\cite{Schuld2021Supervised,Havlek2019Supervised},
promising access to feature spaces that may be hard to reproduce
classically~\cite{Liu2021rigorous}.
Two results have tempered this promise: expressive encoding circuits can
exponentially concentrate their Gram matrices, rendering off-diagonal entries
uninformative~\cite{Thanasilp2024Exponential}, and bandwidth-tuned quantum kernels on
standard benchmarks often collapse to classical RBF~\cite{Shaydulin2022Importance,
FlrezAblan2025similarity}, with dequantization results reinforcing this for broad circuit
classes~\cite{Shin2024Dequantizing, Sahebi2025Dequantization}.
The prevailing view is that quantum kernels offer no practical advantage
on classical data~\cite{Schnabel2025quantum,Kbler2021Inductive,Thanasilp2024Exponential,Liu2021rigorous, Huang2021Power}.
This paper pursues a question complementary to the computational one, an
architectural one: whether a circuit layout that encodes domain-specific
inductive bias can yield finite-sample gains on a structured classical task.
A positive, mechanistically understood instance would be a gateway toward
application-driven quantum-kernel design.

Chaotic time-series forecasting provides a concrete instance of the
structure--similarity mismatch that Euclidean kernels cannot resolve.
The Lorenz-63 attractor~\cite{Lorenz1963Deterministic} is a folded manifold of fractal
dimension ${\approx}\,2.06$ embedded in $\mathbb{R}^3$.
Takens-style delay embedding~\cite{Takens1981Detecting} guarantees that a sufficiently
long window of observations contains the information needed to reconstruct
attractor state, but delay embedding does not imply that the Euclidean metric
on windows is task-optimal: the attractor folds back on itself, so two windows
close in $\|\cdot\|_2$ can lie on different branches of the fold with
divergent futures.
RBF and Mat\'{e}rn kernels~\cite{Rasmussen2005Gaussian} are isotropic functions of
Euclidean distance, and no single bandwidth simultaneously keeps
fold-adjacent, future-divergent windows apart and dynamically similar windows
close.

\paragraph{Contributions}
\begin{itemize}
  \item[\textbf{(C1)}] \AeRot{}, a structured quantum kernel for delay-window
    forecasting that fuses amplitude encoding with grouped angle encoding,
    connecting circuit layout to data temporal structure~\cite{Takens1981Detecting};
    design rules (grouped assignment, no entangling, $\ell^2$ normalization, qubit count)
    are motivated and ablated.
  \item[\textbf{(C2)}] A horizon-resolved benchmark against CV-tuned RBF
    and Mat\'{e}rn baselines over 100 seeds, with crossover at
    ${\sim}0.15$\,tu and $+0.137$ mean $R^2$ gap at $0.25$\,tu (5 qubits). Kernel diagnostics confirm Gram matrices structurally distinct from RBF and strictly higher target-kernel alignment at every horizon. 
    Figure~\ref{fig:overview}[a] is an illustration of the prediction task and quantum-kernel advantage on a representative window.
  \item[\textbf{(C3)}] To our knowledge, the first fair-baseline demonstration of a quantum kernel outperforming tuned classical kernels at forecasting a classical dynamical system, with a mechanistic explanation of where and why the advantage arises. Tail-window Jacobian analysis localises it to the saddle-approach band, where fold-branch ambiguity makes Euclidean similarity structurally insufficient.
\end{itemize}

\section{Related Work}
\label{sec:related}

\paragraph{Kernel methods and classical baselines for chaotic forecasting.}
Tuned RBF kernels and Gaussian processes are the standard fixed-window baseline
for Lorenz-63 regression~\cite{Vlachas2018Data, Shahi2022Prediction}.
Reservoir computing~\cite{Pathak2018Model, Vlachas2019Backpropagation, Gauthier2021Next} and
next-generation RC (NG-RC)~\cite{Gauthier2021Next} dominate long-horizon autonomous
forecasting on Lorenz-63 and Lorenz-96; LSTMs~\cite{Vlachas2018Data} are the
standard recurrent deep-learning baseline.
Takens' delay-embedding theorem~\cite{Takens1981Detecting} is the theoretical scaffold
for any window-based predictor: a delay coordinate of dimension $\geq 2d+1$
recovers the attractor up to diffeomorphism, but does not imply that the
Euclidean metric on windows is task-optimal.
Our paper sits in the fixed-window KRR setting; we restrict our advantage claim
to this regime and treat recurrent methods as context in discussions ( Section~\ref{sec:discussion}  and
Appendix~\ref{sec:esn}).

\paragraph{Quantum kernels for a general ML audience.}

Any variational circuit with a single final measurement is a kernel
method~\cite{Schuld2021Supervised}, making the encoding unitary the
central design decision.  Two pathologies limit current quantum
kernels: expressive circuits concentrate their Gram matrices
exponentially~\cite{Thanasilp2024Exponential}, and bandwidth-tuned
quantum kernels collapse toward classical
RBF~\cite{Shaydulin2022Importance,FlrezAblan2025similarity,Schnabel2025quantum};
dequantization results reinforce this for broad circuit
classes~\cite{Shin2024Dequantizing,Sahebi2025Dequantization,Jerbi2024Shadows}.
The constructive implication~\cite{Thanasilp2024Exponential} is that
geometry-aware embeddings with limited entanglement are the regime
where useful quantum kernels can survive; our work is a concrete
instance of this program.

\paragraph{Encoding strategies.}
Amplitude encoding loads a length-$N = 2^n$ window into the amplitudes of an
$n$-qubit quantum state as an $\ell^2$-normalized superposition; angle encoding
drives single-qubit rotations with data-scaled angles and is data-reuploadable,
giving a universal function approximator with sufficient
repetitions~\cite{PrezSalinas2020Data}.
The two are usually treated as mutually exclusive~\cite{Tudisco2026Evaluating,
Munikote2024Comparing}.
Li et al.~\cite{li2025repetitive} propose repetitive amplitude encoding across multiple
qubit blocks to introduce nonlinearity via inter-block entanglement.
A handful of works combine amplitude and angle encoding for specific
applications~\cite{Chen2025Hybrid, Cowlessur2025QubitEfficient}, but to our knowledge
no prior work has studied this fusion as a kernel.
By encoding multiple data values into each qubit through consecutive rotations,
we exploit data temporal relations and identify per-qubit continuity as the
load-bearing mechanism.

\paragraph{Quantum approaches to time-series and chaos.}
Quantum reservoir computing on Lorenz~\cite{Ahmed2024Prediction, Connerty2024Predicting} uses
recurrence dynamics rather than kernels; it is a complementary mechanism and
not a direct competitor to fixed-window KRR.
QuaCK-TSF~\cite{Aaraba2024QuaCKTSF} is a sliding-window quantum-kernel Gaussian
process applied to synthetic non-chaotic data.
A kernel-ridge formulation of quantum reservoirs and extreme learning
machines has been applied to Lorenz-63 and Mackey--Glass~\cite{Gross2026Kernel},
but it reports intra-quantum efficiency rather than a comparison against
tuned classical baselines.
We are unaware of any prior work that localises a quantum-kernel advantage to a
specific dynamical regime of the underlying system.

\section{The \AeRot{} Quantum Kernel}
\label{sec:kernel}

\paragraph{Notation and data.}
A time-series window $x \in \mathbb{R}^N$ of length $N = 2^n$ from a
single Lorenz channel; the regression target is
$Y_t \in \mathbb{R}^3$ at horizon $h=n_{\rm ahead}\Delta t$.  Each window is per-seed
$z$-scored and $\ell^2$-normalized before encoding.  The qubit count
is bounded by the Lyapunov window rule $N \cdot \Delta t \lesssim
\tau_L \approx 1.1$\,tu, giving $n=4$ ($\Delta t = 0.05$) and
$n=5$ ($\Delta t = 0.025$), both spanning $0.8$\,tu; here $\Delta t = 0.025$
is the RK4 integration step and $n=4$ decimates it by two.  The rule caps the
span, and with it the qubit count at a given sampling interval; the sampling
interval itself is a property of the series rather than a kernel
hyperparameter, and we evaluate both configurations.  Full integration and data-split
details are in Appendix~\ref{app:details}.

\paragraph{Circuit definition.}
Figure.~\ref{fig:overview} (b) and (c) provide schematics of the quantum circuit for kernel \AeRot{}.
The unitary $U(x) = \mathrm{Rot}(\tilde{x}) \cdot A(\tilde{x})$ acts on
$|0\rangle^{\otimes n}$, where $\tilde{x} = x/\|x\|_2$ is the
$\ell^2$-normalized window.
$A(\tilde{x})$ is amplitude encoding: $A(\tilde{x})|0\rangle^{\otimes n} =
\sum_{j=0}^{N-1}\tilde{x}_j|j\rangle$.
$\mathrm{Rot}(\tilde{x})$ is the grouped rotation layer, a tensor product of
independent single-qubit unitaries
$\mathrm{Rot}(\tilde{x}) = \bigotimes_{q=0}^{n-1} W_q(\tilde{x})$,
where qubit $q$ applies $k = \lceil N/n \rceil$ alternating $R_y/R_z$ gates
driven by its assigned contiguous data block:
$W_q(\tilde{x}) = \prod_{m=0}^{k-1}(R_y \text{ or } R_z)(\tilde{x}_{qk+m}/R)$, as illustrated in Figure.~\ref{fig:overview} (c) 
The kernel is
\[
  \kappa(x, y) \;=\;
  \bigl|\langle 0^n | U^\dagger(y)\, U(x) | 0^n \rangle\bigr|^2,
\]
which is PSD by construction and satisfies $\kappa(x,x) = 1$.
Because $\mathrm{Rot} = \bigotimes_q W_q$, the rotation layer introduces
\emph{no entanglement} between qubits: all cross-qubit correlations in
$\kappa$ originate exclusively from the amplitude encoding stage.  As a
consequence, the full kernel admits exact classical evaluation at any qubit
count (Section~\ref{sec:diagnostics}).

\paragraph{Multi-channel fusion.}
Separate kernels $K_x$, $K_y$, $K_z$ are built for each Lorenz channel
and combined as
$K_{\rm fused} = \beta_x K_x + \beta_y K_y + \beta_z K_z$
with $\beta_i \ge 0$, $\sum_i \beta_i = 1$.
The fused kernel is used in KRR to predict all three
output coordinates; the weights $\beta$ are selected by inner CV.

\paragraph{Design motivation.}
Amplitude encoding alone maps each window to a point on $S^{N-1}$: the
resulting kernel $\kappa_{\rm AE}(x,y) = (\tilde{x}\cdot\tilde{y})^2$
is squared cosine similarity, computable classically, and underperforms
tuned RBF in our sweeps.  The rotation layer adds per-qubit local structure
that encodes the \emph{shape} of each temporal block: after $\ell^2$
normalization all windows share the same norm and the rotation gates
resolve within-qubit micro-dynamics that cosine similarity discards.
Together the two stages are complementary: amplitude encoding provides a
globally entangled reference state; the grouped rotation layer applies
bandwidth-tuned, block-local operations on top.  The grouped temporal
assignment mirrors a Takens-style delay
embedding~\cite{Takens1981Detecting}: each qubit encodes a contiguous
local segment, preserving temporal continuity within each qubit.
Per-window $\ell^2$ normalization encodes trajectory shape rather than
absolute position (Appendix~\ref{app:l2norm}); Non-contiguous arrangement or adding entangling layers degrade
the grouped layout by scrambling the structured angle pattern
(Section~\ref{sec:ablations}).

\vspace{-0.7em}
\paragraph{Bandwidth and CV protocol.}

The rotation angle scale $R$ is the quantum analogue of the RBF length
scale~\cite{Shaydulin2022Importance}. The CV-selected optimum $R^\star$ is
chosen per-seed by 5-fold inner CV from a grid centred on $\Rini$; the
RBF bandwidth $\gamma$ and Mat\'{e}rn length scale $\ell$ are tuned on
the same fold structure.  A second inner CV loop selects the Tikhonov
regulariser $\alpha$ and three-channel fusion weights $\beta$.  Grid
specifications are in Appendix~\ref{app:details}.  This protocol
ensures the quantum-vs-classical comparison is not a tuned-vs-untuned
artifact~\cite{Schnabel2025quantum}.

\vspace{-0.5em}
\section{Benchmarks: \AeRot{} vs.\ Tuned Classical Kernels}
\label{sec:benchmarks}
\vspace{-0.5em}
\paragraph{Protocol.}

\begin{figure}[h]
  \centering
   \includegraphics[width=\linewidth]{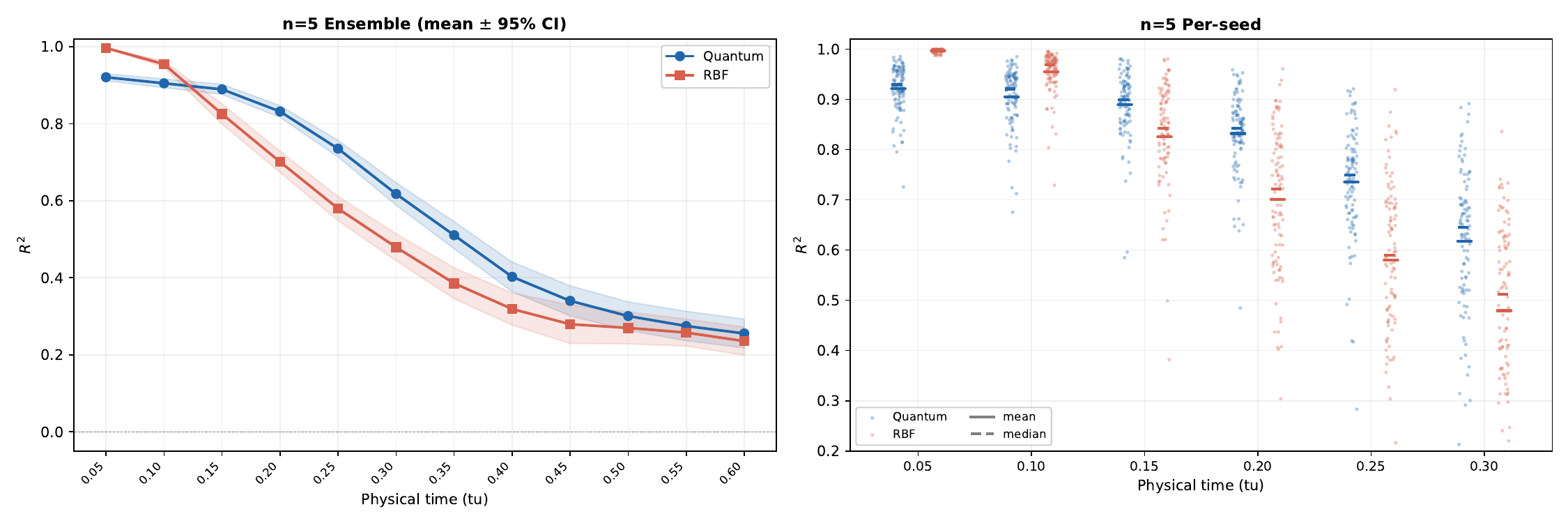}
  \caption{Mean $R^2$ vs.\ prediction horizon for \AeRot{} and tuned RBF
    on $n=5$ (100 seeds, 95\% CI). Left: ensemble mean with shaded
    95\% confidence interval. Right: per-seed scatter plus mean (solid) and median (dashed)
    markers. The \AeRot{} advantage emerges at $0.15$\,tu ($\approx 14\%$ of $\TL$), widens to
    maximum at $0.25$\,tu, then narrows; by $0.60$\,tu collapses as both kernels degrade
    together.}
  \label{fig:horizon}
  \vspace{-1.15em}
\end{figure}

We benchmark on single-step Lorenz-63 KRR across 100 independent trajectory seeds.
For each seed, nested CV selects all hyperparameters (encoding
scale, regulariser, fusion weights) on training data only; the 30-window test set is evaluated exactly once.
Metric: per-seed vector-norm $R^2$ and per-window RMSE across all three Lorenz channels, reported as mean $\pm$ std across 100 seeds. We test $n=5$ (window $N=32$, $\mathrm{d}t=0.025$)
and $n=4$ ($N=16$, $\mathrm{d}t_{\rm eff}=0.05$, results shown in Appendix), chosen so that physical prediction times align.
\vspace{-0.5em}
\paragraph{Main result.}
Table~\ref{tab:main} and Figure~\ref{fig:horizon} shows the $n=5$ results.
At short horizons ($\leq 0.10$\,tu) RBF dominates, as the prediction task is
nearly linear and a smooth Euclidean kernel suffices.
At $0.15$\,tu the advantage reverses: \AeRot{} leads RBF by $+0.050$ in mean
$R^2$.
The gap widens monotonically through $0.25$\,tu ($+0.137$), with $\geq 80\%$ of
seeds favoring \AeRot{} at every horizon past the crossover.
Mat\'{e}rn-5/2 tracks RBF closely, developing a small consistent advantage at
post-crossover horizons ($+0.018$ at $0.25$\,tu), but remains well below
\AeRot{}.
The advantage persists beyond the tabulated range: extended-horizon evaluation
Figure~\ref{fig:horizon} (and Appendix~\ref{app:extended}) shows it remains positive through tested horizon $\approx 0.45$\,tu ($\approx 41\%$ of $\TL$).
The gains concentrate where the dynamics is hard (Section~\ref{sec:regime}).

\begin{table}[t]
  \caption{Mean $R^2 \pm$ standard deviation across 100 seeds for \AeRot{},
    tuned RBF, and Mat\'{e}rn-5/2 on single-step Lorenz-63 KRR ($n=5$,
    $N=32$, $\Ntrain=80$). Gap = \AeRot{} $-$ max(RBF, Mat\'{e}rn). The
    quantum-kernel advantage is positive from $0.15$\,tu and widens monotonically.}
  \label{tab:main}
  \centering
  \small
  \begin{tabular}{r r c c c r}
    \toprule
    $ n\_ahead$ & Phys.\ time & \AeRot{} $R^2$ & RBF $R^2$ &
      Mat\'{e}rn $R^2$ & Gap \\
    \midrule
    2  & 0.05\,tu & $0.921 \pm 0.047$ & $0.997 \pm 0.003$ &
      $0.993 \pm 0.006$ & $-0.076$ \\
    4  & 0.10\,tu & $0.905 \pm 0.058$ & $0.955 \pm 0.042$ &
      $0.950 \pm 0.034$ & $-0.050$ \\
    6  & 0.15\,tu & $0.889 \pm 0.067$ & $0.825 \pm 0.133$ &
      $0.840 \pm 0.101$ & $+0.050$ \\
    8  & 0.20\,tu & $0.832 \pm 0.079$ & $0.701 \pm 0.140$ &
      $0.714 \pm 0.137$ & $+0.118$ \\
    10 & 0.25\,tu & $0.735 \pm 0.110$ & $0.580 \pm 0.160$ &
      $0.598 \pm 0.152$ & $+0.137$ \\
    \bottomrule
  \end{tabular}
  \vspace{-0.8em}
\end{table}

\paragraph{Horizon-resolved signature and the Lyapunov crossover.}

The crossover at $0.15$\,tu ($\approx 14\%$ of $\TL \approx 1.1$\,tu) is an empirical threshold, not a derived one.
Below it, prediction targets of fold-adjacent windows have not yet diverged enough for kernel choice to matter ($R^2 > 0.9$ for all three kernels).
Above it, the fold-resolving inductive bias of \AeRot{} becomes operative and the gap widens monotonically through $\approx 0.45$\,tu ($\approx 41\%$ of $\TL$) before both kernels degrade together (Appendix~\ref{app:extended}).
Both kernels adapt their operating points with horizon \AeRot{} toward smaller $R$, RBF toward larger $\gamma$ reflecting the same bias--variance shift toward finer local resolution (Appendix~\ref{app:hpshift}).
This three-phase pattern (no advantage, growing advantage, joint collapse) is the empirical fingerprint of the mechanism analysed in Section~\ref{sec:regime}.

\section{Where the Advantage Lives: Dynamical Regime Analysis}
\label{sec:regime}
\vspace{-0.5em}
\paragraph{What characterises the hardness?}
The kernel evaluates similarity between length-$W$ delay windows and makes
a prediction $n_{\rm ahead}$ steps beyond the last observed point.
Prediction difficulty is controlled by the local dynamics of the trajectory
\emph{just before} $t_{\rm end}$: the window's early history shapes the
current state, but it is the dynamics in the final steps that determine
whether the prediction target is reachable from the window's endpoint.
We quantify this by the tail-window mean of the largest real Jacobian
eigenvalue,
\[
  \bar\lambda_{\rm tail}^{(w)}(n_{\rm tail})
  \;=\;
  \frac{1}{n_{\rm tail}}
  \sum_{k\,=\,t_{\rm end}-n_{\rm tail}+1}^{t_{\rm end}}
  \max_i\,\mathrm{Re}\!\left(\lambda_i\!\left(J(x_k, y_k, z_k)\right)\right),
\]
where $J$ is the Lorenz Jacobian evaluated at each raw trajectory point.
$\bar\lambda_{\rm tail} > 0$ where the trajectory locally diverges (unstable
tail); $\bar\lambda_{\rm tail} < 0$ where it locally converges (stable tail).
This is a property of the underlying flow only and carries no dependence
on any kernel.
Reference values: origin saddle $\lambda_{\rm max} = +11.83$ (strongly
unstable); lobe centres $C^\pm$, $\lambda_{\rm max} = +0.09$ (weakly
unstable); along the attractor, $\bar\lambda_{\rm tail}$ ranges empirically
from $\sim -3$ (deep in a lobe) to $\sim +8$ (near the saddle).

For each horizon $n_{\rm ahead}$ we sweep $n_{\rm tail} \in
\{1,2,4,8,16,32\}$ and identify the optimal tail length $n_{\rm tail}^*$
that jointly maximises the Pearson correlation between $\bar\lambda_{\rm
tail}$ and per-window RMSE for both kernels.
The $r$-versus-$n_{\rm tail}$ curves have the same shape for both kernels:
rising from the right (long tails dilute the local-instability signal),
peaking at $n_{\rm tail}^*$, and falling on the left (single-point estimates
of $\lambda_{\rm max}$ are noisy).
At the peak, $\bar\lambda_{\rm tail}$ explains $30$--$43\%$ of per-window
variance in either kernel's RMSE (Table~\ref{tab:tail_peak}).
Tail instability is therefore a strong predictor of prediction difficulty for
\emph{both} kernels equally. It is a property of the forecasting task, not
of either kernel's particular weakness.

\begin{figure}[h]
  \centering
  \includegraphics[width=0.8\textwidth]{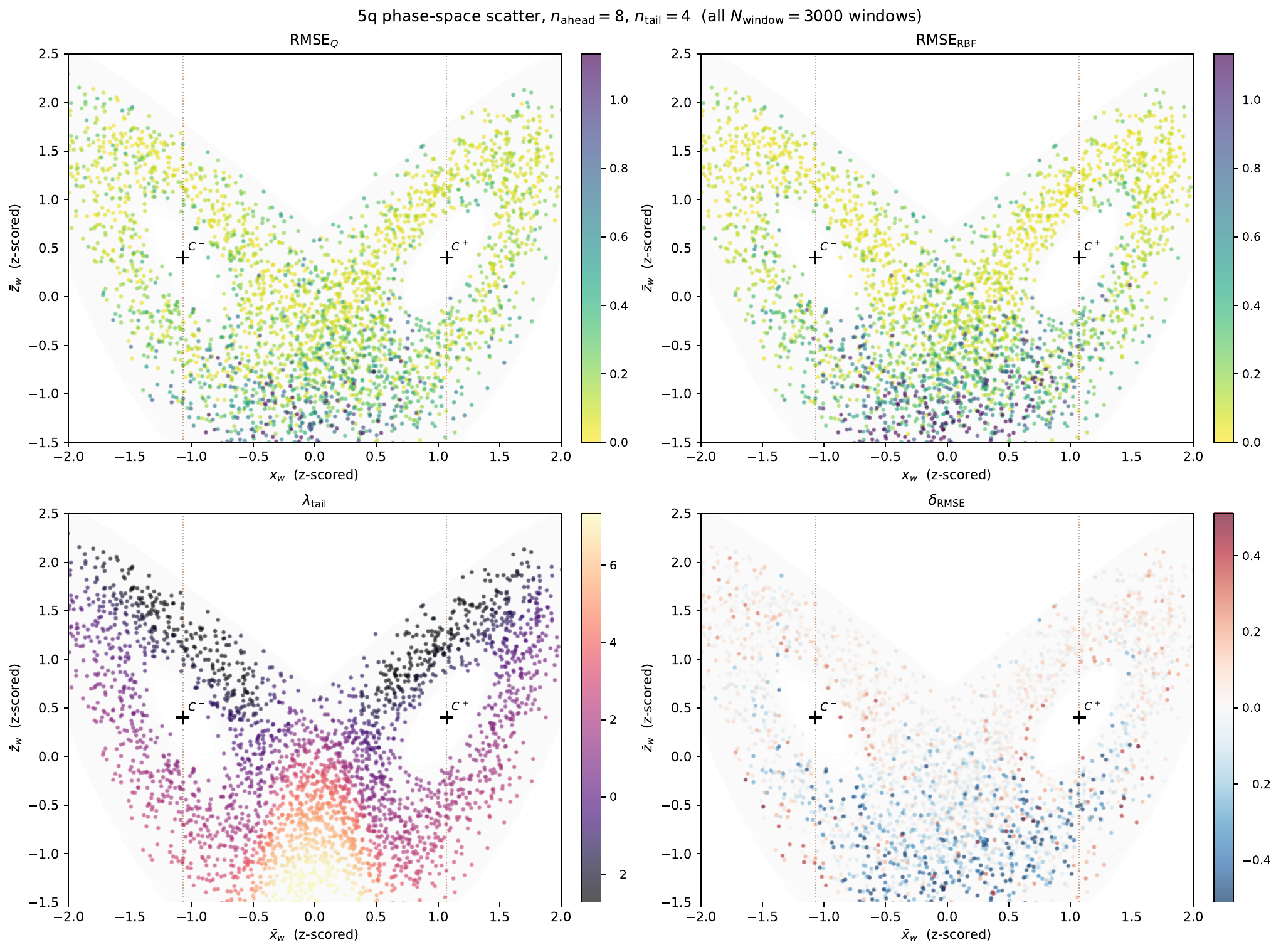}
  \caption{Phase-space scatter of 
  all 3000 windows from 100 seeds.
  $n_{\rm ahead}=8$ ($0.20$\,tu), plotted at tail-mean position
    $(\bar{x}_{\rm tail}, \bar{z}_{\rm tail})$ with $n_{\rm tail}=4$.
    Background: 200k-step reference trajectory.
    Black crosses: Lorenz fixed points $C^\pm$.
    \emph{Panel 1} ($\mathrm{RMSE}_{\AeRot}$) and \emph{Panel 2}
    ($\mathrm{RMSE}_{\rm RBF}$): dark = hard; both kernels fail in the
    same narrow strip (the saddle-approach band).
    \emph{Panel 3} ($\bar\lambda_{\rm tail}$): high instability
    (yellow) co-localises with the same band.
    \emph{Panel 4} ($\delta_{\rm RMSE}$, blue = \AeRot{} wins): the quantum-kernel
    advantage is spatially co-localised with panels 1--3.
    All four quantities point to the same regime.}
  \label{fig:phase_scatter}
  \vspace{-0.5em}
\end{figure}

\begin{table}[t]
  \small
  \begin{minipage}[t]{0.48\linewidth}
    \centering
    \captionof{table}{Optimal tail length $n_{\rm tail}^*$ and peak
      Pearson correlation $r^*$ between $\bar\lambda_{\rm tail}$ and
      per-window RMSE for \AeRot{} and RBF at four horizons.
      $r^*_{\rm RBF} > r^*_{\AeRot}$ at every horizon, indicating RBF
      is more sensitive to tail instability.}
    \label{tab:tail_peak}
    \vspace{4pt}
    \begin{tabular}{r r r r r}
      \toprule
      $n_{\rm ahead}$ & $h$ & $n_{\rm tail}^*$ &
        $r^*_{\AeRot}$ & $r^*_{\rm RBF}$ \\
      \midrule
       8 & 0.20 & 8 & $+0.613$ & $+0.650$ \\
      10 & 0.25 & 4 & $+0.599$ & $+0.657$ \\
      12 & 0.30 & 1 & $+0.594$ & $+0.658$ \\
      14 & 0.35 & 1 & $+0.555$ & $+0.607$ \\
      \bottomrule
    \end{tabular}
  \end{minipage}%
  \hfill
  \begin{minipage}[t]{0.48\linewidth}
    \centering
    \captionof{table}{Target-kernel alignment $A(K_{\rm fused}, YY^\top)$
      ($n{=}5$, 100 seeds, median).  $A_C$ uses the CV-selected fused RBF
      kernel.  \AeRot{} achieves higher alignment at every horizon, with
      $\geq 99\%$ of seeds in favour.}
    \label{tab:tka}
    \vspace{4pt}
    \begin{tabular}{r r r r r}
      \toprule
      $h$ (tu) & $A_Q$ & $A_C$ & $\Delta A$ &
        \% $A_Q{>}A_C$ \\
      \midrule
      0.15 & 0.314 & 0.241 & $+0.076$ &  99\% \\
      0.20 & 0.289 & 0.233 & $+0.057$ &  99\% \\
      0.25 & 0.266 & 0.222 & $+0.046$ & 100\% \\
      \bottomrule
    \end{tabular}
  \end{minipage}
  \vspace{-0.5em}
\end{table}
Two further properties of the $n_{\rm tail}^*$ sweep deserve emphasis.
First, $n_{\rm tail}^*$ decreases monotonically from $8$ at $n_{\rm ahead}=8$
to $1$ at $n_{\rm ahead} \in \{12, 14\}$: as the prediction horizon grows,
the relevant dynamics window shrinks to the last observed step.
Second, $r^*_{\rm RBF} > r^*_{\AeRot}$ at every $(n_{\rm ahead},
n_{\rm tail})$ pair in the full sweep (Appendix~\ref{app:tailfull}), with
the gap roughly constant in $n_{\rm tail}$, suggesting RBF is consistently more
sensitive to tail instability than \AeRot{}.
This asymmetry is the direct engine of the quantum-kernel advantage, as we will show
in the mechanism paragraph below.
\vspace{-0.75em}
\paragraph{Hardness and advantage share the same locus.}
Where on the attractor do hard instances concentrate?
Figure~\ref{fig:phase_scatter} answers this directly by placing each test
window at its tail-mean position $(\bar{x}_{\rm tail}, \bar{z}_{\rm tail})$
in globally $z$-scored coordinates and coloring by four quantities on a
shared scale.
Reading the panels in sequence tells the whole story.

\emph{Panels 1 and 2 ($\mathrm{RMSE}_{\AeRot}$ and $\mathrm{RMSE}_{\rm
RBF}$):} Dark (high-RMSE) markers concentrate in a narrow horizontal strip
near $x \in [-0.5, 0.5]$, $z \in [-1.5, -0.5]$ in globally $z$-scored
coordinates, which is the spatial projection of the origin saddle's stable
manifold, where trajectories funnel toward the saddle before committing to
a lobe.
We call this the \emph{saddle-approach band}.
Crucially, the dark patch is nearly identical for both kernels: the hard
instances are a property of the Lorenz flow, not of either kernel's
particular failure mode.

\emph{Panel 3 ($\bar\lambda_{\rm tail}$):} High tail-instability values
(yellow) are co-localised with exactly the same band.
This closes the loop on question~(1) visually: the saddle-approach band is
hard because the trajectory tail is locally diverging there, and
$\bar\lambda_{\rm tail}$ is the scalar that captures it.

\emph{Panel 4 ($\delta_{\rm RMSE}$):} The \AeRot{} advantage (blue, $\delta_{\rm RMSE} < 0$)
concentrates in the same saddle-approach band where both kernels are hardest
and $\bar\lambda_{\rm tail}$ is highest.
The easy bulk of the attractor (deep in either lobe) shows
$\delta_{\rm RMSE} \approx 0$.
Questions~(3) and (4) therefore have the same answer: \AeRot{} wins where
the tail is locally unstable, which is where the prediction task is hardest,
which is the saddle-approach band.
The four-panel figure makes this alignment exact.

\vspace{-0.5em}
\paragraph{Quantifying the regime split: two mechanisms, not one gradient.}
The scatter plot identifies the locus; the decile analysis makes it
quantitative and reveals that the stable and unstable regimes are not
two ends of a single trend but qualitatively different mechanisms.

\begin{figure}[h]
  \centering
  \includegraphics[width=0.8\textwidth]{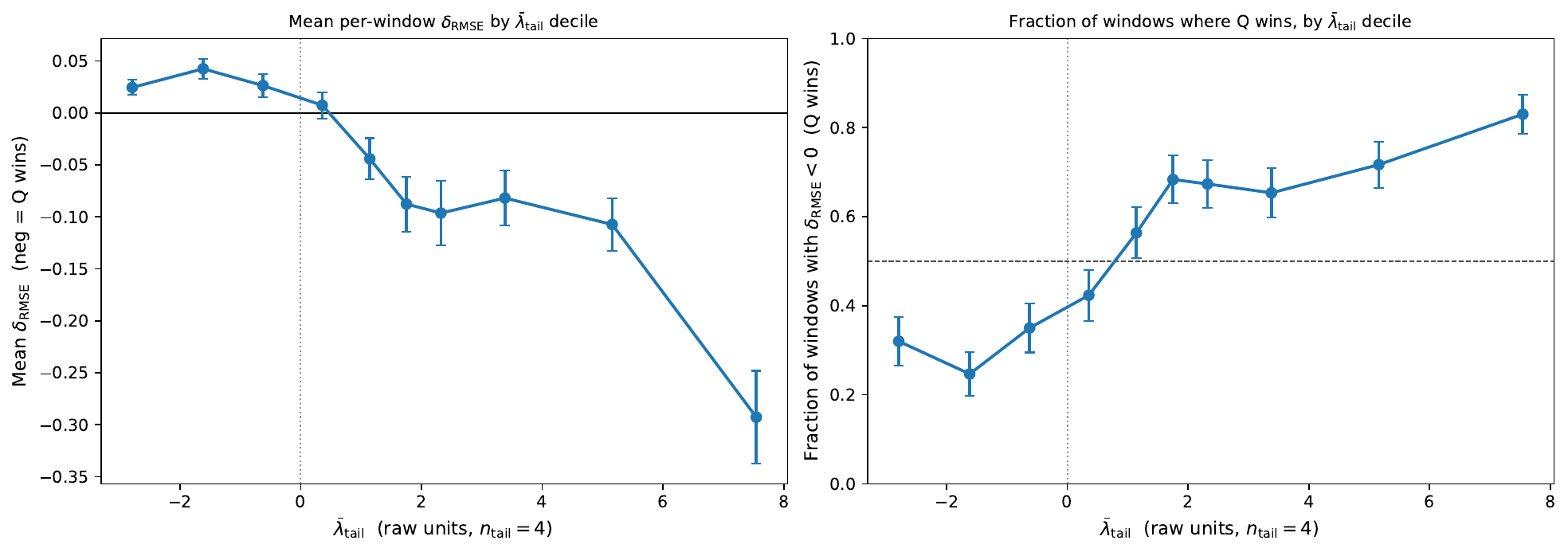}
  \caption{Two-panel decile analysis by $\bar\lambda_{\rm tail}$,
    $n_{\rm ahead}=8$.
    \emph{Left}: mean $\delta_{\rm RMSE} \pm 2$\,SE per decile (negative =
    \AeRot{} wins).
    \emph{Right}: per-decile fraction of windows where \AeRot{} wins
    individually ($\delta_{\rm RMSE} < 0$), $\pm 2$\,SE binomial.
    Vertical dotted line at $\bar\lambda_{\rm tail} = 0$ marks the
    locally stable / locally unstable boundary; horizontal dotted line
    at $0.5$ marks parity.
    The \AeRot{} win rate rises monotonically from $32\%$ at the most
    stable decile to $83\%$ at the most unstable ($\bar\lambda_{\rm
    tail}$).}
  \label{fig:decile_lambda}
  \vspace{-0.5em}
\end{figure}

Binning 3000 pooled windows by $\bar\lambda_{\rm tail}$
(Figure~\ref{fig:decile_lambda}, left panel), the mean $\delta_{\rm RMSE}$ is small
and positive (RBF wins marginally) for the four locally stable deciles
($\bar\lambda_{\rm tail} < 0.4$), crosses zero between decile~4
($\bar\lambda_{\rm tail} \approx +0.36$, $\bar\delta = +0.007$) and
decile~5 ($\bar\lambda_{\rm tail} \approx +1.14$, $\bar\delta = -0.044$),
and grows monotonically more negative through decile~10
($\bar\lambda_{\rm tail} \approx +7.5$, $\bar\delta = -0.29$).
The right panel upgrades the mean statement to a per-window claim: the
fraction of windows where \AeRot{} wins individually rises from
$32\%$ at decile~1 ($\bar\lambda_{\rm tail} \approx -2.8$), crosses $50\%$
near $\bar\lambda_{\rm tail} \approx 0$--$1$, and reaches $\mathbf{83\%}$
at decile~10 ($\pm 4\%$ binomial SE).
\AeRot{} wins almost whenever the tail is strongly unstable.

The regime split is confirmed statistically by subgroup analysis
(Table~\ref{tab:tail_subgroups}).
Splitting by the sign of $\bar\lambda_{\rm tail}$ (stable: $\Nwindow=936$;
unstable: $\Nwindow=2064$):
in the stable subgroup, the correlation between $\bar\lambda_{\rm tail}$
and $\delta_{\rm RMSE}$ is consistent with zero ($p \in \{0.29, 0.83\}$ at the two
horizons); there is no statistically significant advantage mechanism
here, and the mean $\delta_{\rm RMSE} = +0.032$--$+0.036$ reflects a small, fixed
RBF edge independent of how stable the tail is.
In the unstable subgroup, $r(\bar\lambda_{\rm tail}, \delta_{\rm RMSE}) \approx -0.32$
with $p < 10^{-49}$ at both horizons, and the mean $\delta_{\rm RMSE} =
-0.103$--$-0.110$ reflects a substantial, instability-scaling advantage.
Nearly the entire overall correlation ($r = -0.39$,
$p < 10^{-108}$, $N=3000$) is generated by the unstable subgroup alone.

\begin{table}[t]
  \caption{Pearson correlation between $\bar\lambda_{\rm tail}$ and
    per-window $\delta_{\rm RMSE} = \mathrm{RMSE}_{\AeRot} - \mathrm{RMSE}_{\rm
    RBF}$ (negative = \AeRot{} wins), split by sign of
    $\bar\lambda_{\rm tail}$ at $n_{\rm ahead}=8$ ($0.20$\,tu).
    The locally stable subgroup shows no significant correlation ($p=0.29$);
    the unstable subgroup carries essentially all of the overall signal
    ($p < 10^{-10}$).}
  \label{tab:tail_subgroups}
  \centering
  \small
  \begin{tabular}{l r r r r r}
    \toprule
    Subgroup & $\Nwindow$ &
      $r(\bar\lambda_{\rm tail}, \delta_{\rm RMSE})$ & $p$ &
      Mean $\delta_{\rm RMSE}$ & Frac.\ $\delta_{\rm RMSE} < 0$ \\
    \midrule
    Overall                         & 3000 & $-0.393$ & $<10^{-10}$ & $-0.061$ & $54.6\%$ \\
    Stable ($\bar\lambda < 0$)      &  936 & $+0.035$ & $0.29$      & $+0.032$ & $31.2\%$ \\
    Unstable ($\bar\lambda \geq 0$) & 2064 & $-0.317$ & $<10^{-10}$ & $-0.103$ & $65.2\%$ \\
    \bottomrule
  \end{tabular}
  \vspace{-1em}
\end{table}
\vspace{-0.75em}
\paragraph{The mechanism.}
Combining the scatter plot, decile, and subgroup results, two qualitatively
different regimes emerge with two different operative mechanisms. 
In the \textbf{locally stable regime} ($\bar\lambda_{\rm tail} < 0$, both kernels predict accurately, RBF holds a small fixed edge ($\bar\delta \approx +0.03$)  independent of $\bar\lambda_{\rm tail}$, and no fold ambiguity exists to resolve.

In the \textbf{locally unstable regime} ($\bar\lambda_{\rm tail} \geq 0$,
the saddle-approach band), the fold obstruction is active.
The trajectory tail is on the stable manifold of the origin saddle: it is
locally diverging, and two windows that are Euclidean-close in this region
are about to take opposite branches of the Lorenz fold.
$K_{\rm RBF}(x,y) = \exp(-\gamma\|x-y\|^2)$ cannot distinguish these
windows at any bandwidth $\gamma$ because it depends only on $\|x-y\|$,
not on imminent branch identity; it assigns high similarity precisely to
the pairs that are about to diverge most.
\AeRot{}, via its amplitude-encoded $\ell^2$-normalized reference state and
its grouped rotation layer, carries an inductive bias that is not constrained
by Euclidean translation invariance and resolves branch identity better.

The engine of the asymmetry is the consistently higher sensitivity of RBF
to tail instability: $r^*_{\rm RBF} > r^*_{\AeRot}$ at every
$(n_{\rm ahead}, n_{\rm tail})$ pair in the sweep, with the gap roughly
constant.
Both kernels degrade as the tail approaches the saddle, but RBF degrades
faster.
This differential sensitivity directly generates the negative correlation
between $\bar\lambda_{\rm tail}$ and $\delta_{\rm RMSE}$: as the tail becomes more
unstable, both kernels are harder pressed, but RBF more so, widening the
advantage.

The peak advantage is in the \emph{approach} to the saddle, not the crossing itself, where the branch choice is still unresolved. Figure~\ref{fig:overview}[a] gives a concrete single-window example from this regime.

\vspace{-0.5em}
\section{Structural Diagnostics}
\vspace{-0.5em}
\label{sec:diagnostics}

We perform the following diagnostic analysis to establish the kernel
properties that underwrite a finite-sample win, summarised in
Table~\ref{tab:diagmap}: the kernel is discriminative, genuinely distinct
from the tuned classical family, better aligned with the target, and
operating where the geometry gap leaves room for such a gain.  (Because the
kernel is exactly classically evaluable at every size --- closing paragraph
of this section --- these diagnostics do not adjudicate
quantum-versus-classical hardness; they instead characterise the structure of
the kernel itself.)

\begin{table}[h]
  \caption{What each structural diagnostic establishes for the architectural
    claim.}
  \label{tab:diagmap}
  \centering
  \small
  \begin{tabular}{l l}
    \toprule
    Diagnostic & Property established \\
    \midrule
    $\mathrm{Var}_\mathcal{D}[\kappa]$, $\eta_{\rm max}$ &
      discriminative: non-degenerate Gram structure on this data \\
    $\varepsilon_U$, $F$ &
      distinct: not a reparametrisation of the tuned classical family \\
    TKA $A(K, YY^\top)$ &
      aligned: inductive bias matched to the target at every horizon \\
    $g_{CQ}$ &
      bounded finite-$\Ntrain$ room for such a win (classical-vs-classical) \\
    \bottomrule
  \end{tabular}
\end{table}

\paragraph{Gram entry variance and spectral structure.}
At $\Ntrain = 80$ and $0.20$\,tu, the kernel entry variance
$\mathrm{Var}_{\mathcal{D}}[\kappa]$ is $2.24$ (4q) and $1.51$ (5q),
against an RBF reference of $2.18$: both are strictly positive and of the
same order as the classical reference.  For 4q, quantum
and RBF variance are comparable (ratio $1.03$); for 5q, the quantum
spectrum is flatter (ratio $0.69$), consistent with the larger feature
space distributing similarity mass more evenly.  The normalized leading
eigenvalue $\eta_{\max}$ is $0.123$ (4q) and $0.086$ (5q), both well
above the flat-spectrum limit $1/\Ntrain = 0.012$ and far from rank-1
degeneracy.  These are fixed-$(n, \Ntrain)$ observations on the kernel's
discriminative structure.  (Evaluation cost, by contrast, is settled at every
size; see the closing paragraph.)

\vspace{-0.75em}
\paragraph{Expressivity, structural distinctness, and target-kernel alignment.}
The expressivity measure $\varepsilon_U = 0.178$ (4q) and $0.146$ (5q) at
$0.20$\,tu, near the respective large-$R$ limits, indicates that both circuits
operate in the high-expressivity regime~\cite{FlrezAblan2025similarity}.
Combined with $\mathrm{Var}_\mathcal{D} = 2.24$ ($n{=}4$) and $1.51$ ($n{=}5$), the
Gram entries remain informative at these sizes even at large $R$, a departure
from the separable-circuit regime analysed by Fl\'{o}rez-Ablan et
al.~\cite{FlrezAblan2025similarity} that we attribute to the non-separable
amplitude encoding layer.
The normalized Frobenius distance $F = \|K_Q - K_C\|_F / \|K_Q\|_F$ at
$0.20$\,tu is $0.436$ (4q) and $0.462$ (5q), confirming structural distinctness:
the two Gram matrices differ by $44$--$46\%$ of the quantum kernel's own norm,
not a reparametrisation.
Target-kernel alignment (TKA) $A(K, YY^\top)$~\cite{Cristianini2001Kernel} is higher
for \AeRot{} than for RBF at every tested horizon and in both configurations
(Table~\ref{tab:tka}), with $95$--$100\%$ of seeds favoring \AeRot{} at every
(config, horizon) pair.
The alignment gap narrows monotonically with horizon but remains positive
throughout, providing a task-specific account of the empirical $R^2$ advantage.

\vspace{-0.75em}
\paragraph{Classical evaluability and relation to dequantization.}
The scope of this claim deserves precision, and it is sharper than the usual
``we do not claim non-dequantizability.''  Because the rotation layer is a
tensor product of single-qubit unitaries, the \emph{full} \AeRot{} kernel
--- not merely its amplitude-encoding stage --- admits exact classical
evaluation: $\phi(x) = \mathrm{Rot}(\tilde{x})\,A(\tilde{x})\,
|0^n\rangle$ is an explicit $N$-vector computable in $O(N \log N)$ once per
window, and each Gram entry $\kappa(x,y) = |\langle \phi(x),
\phi(y)\rangle|^2$ costs $O(N)$.  Since the input is the raw length-$N$
window, this matches the cost any evaluator --- classical, or quantum through
state preparation --- must already pay to read or load the data.  No
evaluation-side quantum speedup therefore exists at any qubit count, and
every number in this paper is an exact classical computation.  The
contribution is architectural: a quantum-circuit-derived similarity whose
inductive bias outperforms standard classical kernels at matched tuning, in
the bounded finite-$\Ntrain$ regime permitted by the geometric-difference
framework~\cite{Huang2021Power}.  Consistently, the geometric difference at
the primary horizon ($0.20$\,tu), computed in the advantage-relevant
direction $g_{CQ}$ (which inverts the \emph{classical}
kernel~\cite{Huang2021Power}), is $g_{CQ} \approx 2.3$ (4q) and
$\approx 3.6$ (5q), well below $\sqrt{\Ntrain} \approx 8.94$.  At the tested sizes, $g > 1$
leaves room for the bounded finite-sample advantage that our $R^2$ and
alignment results occupy.  These statements are specific to the
entanglement-free rotation layer of this paper (the $\theta = 0$ point of
the entangling sweep in Section~\ref{sec:ablations}); kernels in this family
that activate entangling layers are outside the scope of this argument, and
we make no claim here about their dequantization status.

\section{Architectural Ablations}
\label{sec:ablations}

\begin{figure}[!h]
  \centering
   \includegraphics[width=\linewidth]{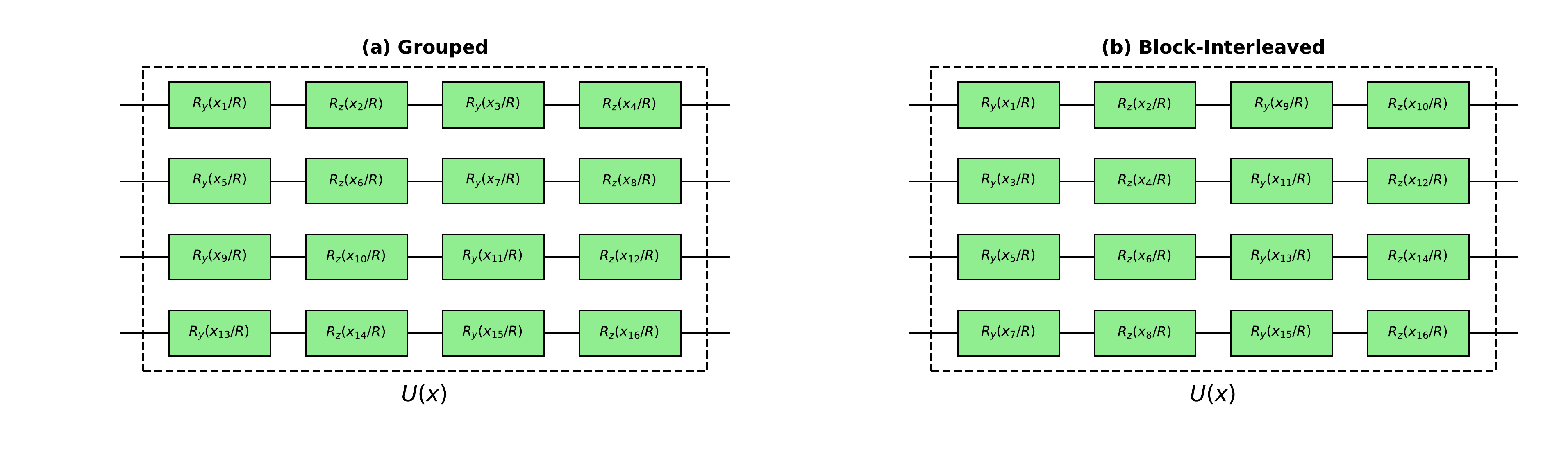}
  \caption{Angle-assignment layouts for the rotation layer. $n=4$ qubits, $N=16$. 
  (a)  grouped: each qubit receives a contiguous 4-step block. (b)
    block-interleaved: each qubit receives 2 steps from the first half
    and 2 from the second half of the window, breaking temporal continuity.}
  \label{fig:group_vs_interleaved}
  \vspace{-0.5em}
\end{figure}

The critical design choice is how the $2^n$ window angles are assigned to $n$
qubits.
In the \emph{grouped} (default) layout, each qubit owns a contiguous slice:
qubit $q$ receives steps $\{qk+1, \ldots, qk+k\}$ with $k = \lceil N/n\rceil$.
In the \emph{block-interleaved} layout, angles are consumed in qubit-major
round-robin order across $M_{\rm rot}=2$ blocks, giving each qubit two
non-adjacent temporal segments.
Figure~\ref{fig:group_vs_interleaved} illustrates both layouts for $n=4$, $N=16$.

Entangling gates are a standard tool in quantum machine learning circuits
for generating entanglement between qubits and enabling exploration of a
broader region of Hilbert space~\cite{Sim2019Expressibility}; we study
their effect in our setup by sweeping a brick-wall IsingXX($\theta$)
entangling layer from $\theta=0$ (identity) to $\theta=\pi/2$ (locally
equivalent to CNOT).
The two layouts behave oppositely (Appendix~\ref{App:group_ablation}): adding
entanglement \emph{monotonically degrades} the grouped kernel (mean $R^2$
from $0.858$ at $\theta=0$ to $0.813$ at CNOT), while it \emph{improves}
the interleaved kernel ($0.775$ to $0.788$).
Despite this, the interleaved layout never surpasses grouped at any entangling
strength.

The mechanism is temporal locality: the grouped layout preserves data temporal
continuity encoded in each qubit, which is the Takens-compatible inductive
bias.
Entanglement partially recovers cross-temporal information in the interleaved
case by stitching together its scattered segments, but cannot reconstruct the
contiguous structure that grouped supplies without any entanglement.
For the grouped layout, entangling layers introduce cross-block interference
that destroys the structured temporal pattern each qubit accumulates,
explaining the monotonic degradation.
Temporal locality is the primary inductive bias; entanglement is a partial
remedy for interleaved, and a degradation for grouped.

\section{Discussion and Conclusion}
\label{sec:discussion}

We introduced \AeRot{}, a quantum kernel fusing amplitude encoding
with grouped angle encoding, and benchmarked it against tuned
classical kernels on Lorenz-63 KRR across 100 seeds.  The kernel
outperforms tuned RBF and Mat\'ern-5/2 from $0.15$\,tu, with the
gap widening to $+0.137$ mean~$R^2$ at $0.25$\,tu.  Tail-window
Jacobian analysis localises the advantage to the saddle-approach
band, where fold-branch ambiguity makes Euclidean similarity
structurally limited.

The advantage is architectural, not complexity-theoretic: the kernel itself
is exactly classically evaluable at any size (Section~\ref{sec:diagnostics}),
the geometric difference $g_{CQ}$ sits well below $\sqrt{\Ntrain}$ at
the primary horizon, and the gain is
better characterised as an inductive-bias match between the encoding
and the attractor's fold structure, measurable via target-kernel
alignment at every horizon.  K\"ubler et al.~\cite{Kbler2021Inductive}
identify a task-aligned RKHS as the necessary condition for quantum
kernel advantage; our results are consistent with this condition on
classical chaotic data, without appeal to quantum structure in the
data itself.

The prevailing assumption, supported by the only rigorous end-to-end separation resting on a cryptographically constructed classical task~\cite{Liu2021rigorous} and by quantum kernel benchmarks focusing on quantum or synthetic datasets~\cite{Huang2021Power}, is that quantum kernels need engineered or quantum structure in the data to outperform classical alternatives.
Our result shows that geometric structure in a classical system, specifically a folded attractor where Euclidean distance is misaligned with dynamical similarity, can satisfy the RKHS alignment condition through circuit architecture alone: the relevant criterion is not whether the data is quantum, but whether the task has geometric obstructions that isotropic kernels cannot resolve at finite size.

The regime analysis distinguishes this work from a benchmark comparison.  The advantage is not only measured but localised and predicted: given a test window's tail-instability diagnostic $\bar\lambda_{\rm tail}$, one can predict which kernel wins on that window with 83\% accuracy in the most unstable decile, and the sign-flip near $\bar\lambda_{\rm tail} \approx 0$ provides a
physics-grounded decision boundary. This transforms the result from ``the quantum kernel is better on average'' to a per-instance, mechanistically explained advantage tied to a computable property of the underlying flow. 

The tail-window Jacobian analysis is not specific to Lorenz-63. Any chaotic system has local stability structure characterisable by Jacobian eigenvalues, and the question ``does the quantum-kernel advantage concentrate in dynamically unstable regions?''\ can be posed for Lorenz-96, Kuramoto--Sivashinsky, or turbulent flows with richer fold geometry. The diagnostic pipeline, local Jacobian spectrum, decile binning, stable/unstable subgroup split, transfers directly; Lorenz-63 serves as the first instance of this analysis framework rather than its endpoint.

\paragraph{Limitations and future work.}
(i)~Amplitude encoding in \AeRot{} requires $O(2^n)$ gates for arbitrary
states; hardware implementations should target structured inputs
where this cost is reduced.  NISQ noise is expected to blunt the
advantage below its idealised value.
(ii)~The advantage operates in a middle-horizon band
(${\approx}\,0.14$--$0.41\,\TL$); beyond $0.45$\,tu both kernels
degrade and the gap collapses. At longer horizons the prediction likely depends on joint structure across all three Lorenz channels, but 
CV consistently selects near-unity weight on the  $y$-channel alone ($\beta_y \approx 0.9$); encoding multiple channels into the circuit jointly rather than 
fusing separate single-channel kernels is a natural 
next step.
(iii)~Our findings are specific to Lorenz-63; extension to
higher-dimensional chaos (Lorenz-96, Kuramoto--Sivashinsky) is
needed to test whether the saddle-approach advantage generalises
to systems with richer fold geometry.  Because the window rule caps the span from
the system's memory timescale (Section~\ref{sec:kernel}), each system and
sampling rate bounds its own qubit count; whether the grouped layout retains
its edge when each qubit must absorb more angles is untested.
(iv)~Echo state networks substantially outperform all fixed-window
methods ($R^2 > 0.95$ at $0.25$\,tu; Appendix~\ref{sec:esn}),
reflecting access to full trajectory history rather than a fixed
$N$-step window.  Our advantage claim is confined to the
fixed-window KRR regime; whether a quantum recurrent architecture
can close the gap with ESNs is an open question.
(v)~The claim of this paper is a quantum-kernel win at the training size
and window length studied, in the fixed-window setting.

\begin{ack}
  This work was supported by the National Aeronautics and Space
  Administration under Award No.\ 80MSFC25M0084.  Z.~Wang and A.~Akbari are
  supported by NASA funding through cooperative agreement 80NSSC24M0035.
\end{ack}

\newpage
\bibliographystyle{unsrtnat}
\bibliography{kernel}

\begin{thebibliography}{34}
\providecommand{\natexlab}[1]{#1}
\providecommand{\url}[1]{\texttt{#1}}
\expandafter\ifx\csname urlstyle\endcsname\relax
  \providecommand{\doi}[1]{doi: #1}\else
  \providecommand{\doi}{doi: \begingroup \urlstyle{rm}\Url}\fi

\bibitem[Schuld(2021)]{Schuld2021Supervised}
Maria Schuld.
\newblock Supervised quantum machine learning models are kernel methods.
\newblock \emph{arXiv}, 2021.
\newblock \doi{10.48550/arxiv.2101.11020}.
\newblock URL \url{https://doi.org/10.48550/arxiv.2101.11020}.

\bibitem[Havlíček et~al.(2019)Havlíček, Córcoles, Temme, Harrow, Kandala,
  Chow, and Gambetta]{Havlek2019Supervised}
Vojtěch Havlíček, Antonio~D. Córcoles, Kristan Temme, Aram~W. Harrow,
  Abhinav Kandala, Jerry~M. Chow, and Jay~M. Gambetta.
\newblock Supervised learning with quantum-enhanced feature spaces.
\newblock \emph{Nature}, 2019.
\newblock \doi{10.1038/s41586-019-0980-2}.
\newblock URL \url{https://doi.org/10.1038/s41586-019-0980-2}.

\bibitem[Liu et~al.(2021)Liu, Arunachalam, and Temme]{Liu2021rigorous}
Yunchao Liu, Srinivasan Arunachalam, and Kristan Temme.
\newblock A rigorous and robust quantum speed-up in supervised machine
  learning.
\newblock \emph{Nature Physics}, 2021.
\newblock \doi{10.1038/s41567-021-01287-z}.
\newblock URL \url{https://doi.org/10.1038/s41567-021-01287-z}.

\bibitem[Thanasilp et~al.(2024)Thanasilp, Wang, Cerezo, and
  Holmes]{Thanasilp2024Exponential}
Supanut Thanasilp, Samson Wang, M.~Cerezo, and Zoë Holmes.
\newblock Exponential concentration in quantum kernel methods.
\newblock \emph{Nature Communications}, 2024.
\newblock \doi{10.1038/s41467-024-49287-w}.
\newblock URL \url{https://doi.org/10.1038/s41467-024-49287-w}.

\bibitem[Shaydulin and Wild(2022)]{Shaydulin2022Importance}
Ruslan Shaydulin and Stefan~M. Wild.
\newblock Importance of kernel bandwidth in quantum machine learning.
\newblock \emph{Physical Review A}, 2022.
\newblock \doi{10.1103/physreva.106.042407}.
\newblock URL \url{https://doi.org/10.1103/physreva.106.042407}.

\bibitem[Flórez-Ablan et~al.(2025)Flórez-Ablan, Roth, and
  Schnabel]{FlrezAblan2025similarity}
Roberto Flórez-Ablan, Marco Roth, and Jan Schnabel.
\newblock On the similarity of bandwidth-tuned quantum kernels and classical
  kernels.
\newblock \emph{Quantum Science and Technology}, 2025.
\newblock \doi{10.1088/2058-9565/ade7ad}.
\newblock URL \url{https://doi.org/10.1088/2058-9565/ade7ad}.

\bibitem[Shin et~al.(2024)Shin, Teo, and Jeong]{Shin2024Dequantizing}
S.~Shin, Yong-Siah Teo, and H.~Jeong.
\newblock Dequantizing quantum machine learning models using tensor networks.
\newblock \emph{Physical Review Research}, 2024.
\newblock \doi{10.1103/physrevresearch.6.023218}.
\newblock URL \url{https://doi.org/10.1103/physrevresearch.6.023218}.

\bibitem[Sahebi et~al.(2025)Sahebi, Barthe, Suzuki, Holmes, and
  Grossi]{Sahebi2025Dequantization}
Mehrad Sahebi, Alice Barthe, Yudai Suzuki, Zoë Holmes, and Michele Grossi.
\newblock On dequantization of supervised quantum machine learning via random
  fourier features.
\newblock \emph{arXiv}, 2025.

\bibitem[Schnabel and Roth(2024)]{Schnabel2025quantum}
Jan Schnabel and Marco Roth.
\newblock Quantum kernel methods under scrutiny: a benchmarking study.
\newblock \emph{arXiv preprint arXiv:2409.04406}, 2024.

\bibitem[Kübler et~al.(2021)Kübler, Buchholz, and
  Schölkopf]{Kbler2021Inductive}
Jonas~M. Kübler, Simon Buchholz, and Bernhard Schölkopf.
\newblock The inductive bias of quantum kernels.
\newblock \emph{Neural Information Processing Systems}, 2021.
\newblock \doi{10.48550/arxiv.2106.03747}.
\newblock URL \url{https://doi.org/10.48550/arxiv.2106.03747}.

\bibitem[Huang et~al.(2021)Huang, Broughton, Mohseni, Babbush, Boixo, Neven,
  and McClean]{Huang2021Power}
Hsin-Yuan Huang, Michael Broughton, Masoud Mohseni, Ryan Babbush, Sergio Boixo,
  Hartmut Neven, and Jarrod~R McClean.
\newblock Power of data in quantum machine learning.
\newblock \emph{Nature communications}, 12\penalty0 (1):\penalty0 2631, 2021.

\bibitem[Lorenz(1963)]{Lorenz1963Deterministic}
Edward~N. Lorenz.
\newblock Deterministic nonperiodic flow.
\newblock \emph{Journal of the Atmospheric Sciences}, 1963.
\newblock \doi{10.1175/1520-0469(1963)020<0130:dnf>2.0.co;2}.
\newblock URL
  \url{https://doi.org/10.1175/1520-0469(1963)020<0130:dnf>2.0.co;2}.

\bibitem[Takens(1981)]{Takens1981Detecting}
Floris Takens.
\newblock Detecting strange attractors in turbulence.
\newblock \emph{Lecture notes in mathematics}, 1981.
\newblock \doi{10.1007/bfb0091924}.
\newblock URL \url{https://doi.org/10.1007/bfb0091924}.

\bibitem[Rasmussen and Williams(2005)]{Rasmussen2005Gaussian}
Carl~Edward Rasmussen and Christopher K.~I. Williams.
\newblock Gaussian processes for machine learning.
\newblock \emph{The MIT Press eBooks}, 2005.
\newblock \doi{10.7551/mitpress/3206.001.0001}.
\newblock URL \url{https://doi.org/10.7551/mitpress/3206.001.0001}.

\bibitem[Vlachas et~al.(2018)Vlachas, Byeon, Wan, Sapsis, and
  Koumoutsakos]{Vlachas2018Data}
Pantelis~R. Vlachas, Wonmin Byeon, Zhong~Y. Wan, Themistoklis~P. Sapsis, and
  Petros Koumoutsakos.
\newblock Data-driven forecasting of high-dimensional chaotic systems with long
  short-term memory networks.
\newblock \emph{Proc. R. Soc. A 474 (2018) 20170844}, 2018.
\newblock URL \url{https://arxiv.org/abs/1802.07486}.

\bibitem[Shahi et~al.(2022)Shahi, Fenton, and Cherry]{Shahi2022Prediction}
S.~Shahi, F.~Fenton, and E.~Cherry.
\newblock Prediction of chaotic time series using recurrent neural networks and
  reservoir computing techniques: A comparative study.
\newblock \emph{Machine Learning with Applications}, 2022.
\newblock \doi{10.1016/j.mlwa.2022.100300}.
\newblock URL \url{https://doi.org/10.1016/j.mlwa.2022.100300}.

\bibitem[Pathak et~al.(2018)Pathak, Hunt, Girvan, Lu, and Ott]{Pathak2018Model}
Jaideep Pathak, Brian~R. Hunt, Michelle Girvan, Zhixin Lu, and Edward Ott.
\newblock Model-free prediction of large spatiotemporally chaotic systems from
  data: A reservoir computing approach.
\newblock \emph{Physical Review Letters}, 2018.
\newblock \doi{10.1103/physrevlett.120.024102}.
\newblock URL \url{https://doi.org/10.1103/physrevlett.120.024102}.

\bibitem[Vlachas et~al.(2019)Vlachas, Pathak, Hunt, Sapsis, Girvan, Ott, and
  Koumoutsakos]{Vlachas2019Backpropagation}
Pantelis~R. Vlachas, Jaideep Pathak, Brian~R. Hunt, Themistoklis~P. Sapsis,
  Michelle Girvan, Edward Ott, and Petros Koumoutsakos.
\newblock Backpropagation algorithms and reservoir computing in recurrent
  neural networks for the forecasting of complex spatiotemporal dynamics.
\newblock \emph{Neural Networks}, 2019.
\newblock \doi{10.1016/j.neunet.2020.02.016}.
\newblock URL \url{https://doi.org/10.1016/j.neunet.2020.02.016}.

\bibitem[Gauthier et~al.(2021)Gauthier, Bollt, Griffith, and
  Barbosa]{Gauthier2021Next}
Daniel~J. Gauthier, Erik Bollt, Aaron Griffith, and Wendson A.~S. Barbosa.
\newblock Next generation reservoir computing.
\newblock \emph{Nature Communications}, 2021.
\newblock \doi{10.1038/s41467-021-25801-2}.
\newblock URL \url{https://doi.org/10.1038/s41467-021-25801-2}.

\bibitem[Jerbi et~al.(2024)Jerbi, Gyurik, Marshall, Molteni, and
  Dunjko]{Jerbi2024Shadows}
Sofiène Jerbi, Casper Gyurik, Simon~C. Marshall, Riccardo Molteni, and Vedran
  Dunjko.
\newblock Shadows of quantum machine learning.
\newblock \emph{Nature Communications}, 2024.
\newblock \doi{10.1038/s41467-024-49877-8}.
\newblock URL \url{https://doi.org/10.1038/s41467-024-49877-8}.

\bibitem[Pérez-Salinas et~al.(2020)Pérez-Salinas, Cervera-Lierta, Gil-Fuster,
  and Latorre]{PrezSalinas2020Data}
Adrián Pérez-Salinas, Alba Cervera-Lierta, Elies Gil-Fuster, and José~I.
  Latorre.
\newblock Data re-uploading for a universal quantum classifier.
\newblock \emph{Quantum}, 2020.
\newblock \doi{10.22331/q-2020-02-06-226}.
\newblock URL \url{https://doi.org/10.22331/q-2020-02-06-226}.

\bibitem[Tudisco(2026)]{Tudisco2026Evaluating}
L.~Tudisco.
\newblock Evaluating angle and amplitude encoding strategies for variational
  quantum machine learning: their impact on model’s accuracy.
\newblock 2026.

\bibitem[Munikote(2024)]{Munikote2024Comparing}
Nidhi Munikote.
\newblock Comparing quantum encoding techniques.
\newblock 2024.
\newblock URL \url{https://arxiv.org/abs/2410.09121}.

\bibitem[Li et~al.(2025)Li, Fu, Meng, and Du]{li2025repetitive}
Ziyang Li, Xiaofei Fu, Lingdong Meng, and Ruishan Du.
\newblock A repetitive amplitude encoding method for enhancing the mapping
  ability of quantum neural networks.
\newblock \emph{Scientific Reports}, 15:\penalty0 32111, 2025.

\bibitem[Chen et~al.(2025)Chen, Griffin, Recchia, Zhou, and
  Zhang]{Chen2025Hybrid}
Ying Chen, Paul Griffin, Paolo Recchia, Lei Zhou, and Hongrui Zhang.
\newblock Hybrid quantum neural networks with amplitude encoding: Advancing
  recovery rate predictions.
\newblock 2025.
\newblock URL \url{https://arxiv.org/abs/2501.15828}.

\bibitem[Cowlessur et~al.(2025)Cowlessur, Alpcan, Thapa, Camtepe, and
  Kundu]{Cowlessur2025QubitEfficient}
Hevish Cowlessur, Tansu Alpcan, Chandra Thapa, Seyit Camtepe, and Neel~Kanth
  Kundu.
\newblock A qubit-efficient hybrid quantum encoding mechanism for quantum
  machine learning.
\newblock 2025.
\newblock URL \url{https://arxiv.org/abs/2506.19275}.

\bibitem[Ahmed et~al.(2024)Ahmed, Tennie, and Magri]{Ahmed2024Prediction}
Osama Ahmed, Felix Tennie, and Luca Magri.
\newblock Prediction of chaotic dynamics and extreme events: A recurrence-free
  quantum reservoir computing approach.
\newblock \emph{Physical Review Research}, 6\penalty0 (4):\penalty0 043082,
  2024.

\bibitem[Connerty et~al.(2024)Connerty, Evans, Angelatos, and
  Narayanan]{Connerty2024Predicting}
Erik Connerty, Ethan~N. Evans, Gerasimos Angelatos, and Vignesh Narayanan.
\newblock Predicting chaotic systems with quantum echo-state networks.
\newblock 2024.
\newblock URL \url{https://arxiv.org/abs/2412.07910}.

\bibitem[Aaraba et~al.(2024)Aaraba, Cherkaoui, Ahmad, Laprade,
  Nahman-Lévesque, Vieloszynski, and Wang]{Aaraba2024QuaCKTSF}
Abdallah Aaraba, Soumaya Cherkaoui, Ola Ahmad, Jean-Frédéric Laprade, Olivier
  Nahman-Lévesque, Alexis Vieloszynski, and Shengrui Wang.
\newblock Quack-tsf: Quantum-classical kernelized time series forecasting.
\newblock 2024.
\newblock URL \url{https://arxiv.org/abs/2408.12007}.

\bibitem[Gross and Rieser(2026)]{Gross2026Kernel}
Markus Gross and Hans-Martin Rieser.
\newblock Kernel-based optimization of measurement operators for quantum
  reservoir computers.
\newblock \emph{arXiv preprint arXiv:2602.14677}, 2026.

\bibitem[Cristianini et~al.(2001)Cristianini, Shawe-Taylor, Elisseeff, and
  Kandola]{Cristianini2001Kernel}
Nello Cristianini, John Shawe-Taylor, André Elisseeff, and Jaz Kandola.
\newblock On kernel-target alignment.
\newblock \emph{Neural Information Processing Systems}, 2001.
\newblock \doi{10.1007/3-540-33486-6_8}.
\newblock URL \url{https://doi.org/10.1007/3-540-33486-6_8}.

\bibitem[Sim et~al.(2019)Sim, Johnson, and
  Aspuru‐Guzik]{Sim2019Expressibility}
Sukin Sim, Peter~D. Johnson, and Alán Aspuru‐Guzik.
\newblock Expressibility and entangling capability of parameterized quantum
  circuits for hybrid quantum‐classical algorithms.
\newblock \emph{Advanced Quantum Technologies}, 2019.
\newblock \doi{10.1002/qute.201900070}.
\newblock URL \url{https://doi.org/10.1002/qute.201900070}.

\bibitem[Jaeger and Haas(2004)]{Jaeger2004Harnessing}
Herbert Jaeger and Harald Haas.
\newblock Harnessing nonlinearity: Predicting chaotic systems and saving energy
  in wireless communication.
\newblock \emph{Science}, 2004.
\newblock \doi{10.1126/science.1091277}.
\newblock URL \url{https://doi.org/10.1126/science.1091277}.

\bibitem[Schuld et~al.(2021)Schuld, Sweke, and Meyer]{Schuld2021effect}
Maria Schuld, Ryan Sweke, and Johannes~Jakob Meyer.
\newblock The effect of data encoding on the expressive power of variational
  quantum-machine-learning models.
\newblock \emph{Physical Review A}, 2021.
\newblock \doi{10.1103/PhysRevA.103.032430}.
\newblock URL \url{https://doi.org/10.1103/PhysRevA.103.032430}.

\end{thebibliography}

\newpage 
\appendix

\section{Extended Related Work: Quantum Kernels}
\label{app:qk_background}

Variational quantum circuits have emerged as a flexible class of models for
supervised learning, capable in principle of representing functions over
exponentially large feature spaces.  A key theoretical result is that any
such circuit with a single final measurement is a kernel method: training
variationally and training via the kernel trick yield the same hypothesis
class~\cite{Schuld2021Supervised}, so the circuit's data-encoding unitary
completely determines what the model can express.  The quantum feature map
embeds a classical input $x$ into a $2^n$-dimensional Hilbert space via a
unitary $U(x)$, and the kernel is the squared overlap
$\kappa(x,y) = |\langle 0^n | U^\dagger(y)\,U(x) | 0^n \rangle|^2$.
This makes the encoding choice, not the trainable parameters, the central
design decision.

Two practical pathologies have tempered early optimism.  First, expressive
encoding circuits can cause off-diagonal kernel entries to concentrate
exponentially in $n$ around a fixed value, making the Gram matrix
information-free~\cite{Thanasilp2024Exponential}.  Thanasilp et
al.\ identify four contributing causes: high circuit expressibility, global
measurements, high entanglement, and hardware noise.  Their constructive
implication is that problem-inspired, geometry-aware embeddings with limited
entanglement are the regime where useful quantum kernels can survive; the
\AeRot{} kernel is designed to operate in exactly this regime (amplitude
encoding provides entanglement through data loading rather than trainable
gates, and the rotation layer is a tensor product with no additional
entangling operations).

Second, even non-concentrating kernels often collapse toward classical RBF
under bandwidth tuning: once the angle scale is cross-validated, quantum
kernels on standard classification benchmarks match but rarely dominate
tuned classical kernels~\cite{Shaydulin2022Importance,
FlrezAblan2025similarity, Schnabel2025quantum}.  Flórez-Ablan et
al.~\cite{FlrezAblan2025similarity} show that for product-state
angle-encoding circuits, varying the bandwidth parameter traces a path in
kernel space that passes through (or near) an RBF kernel, so any advantage
such circuits appear to offer can often be recovered by tuning a classical
RBF bandwidth.  A recent benchmarking study~\cite{Schnabel2025quantum} finds no
configuration that consistently outperforms tuned classical alternatives.  These results apply to
product-state angle encodings on generic classification tasks; \AeRot{}
departs from this regime by using amplitude encoding (which produces an
entangled state for generic inputs) on a structured time-series task.

On the asymptotic side, dequantization results show that broad families of
quantum kernel models admit efficient classical surrogates.  Tensor-network
methods can approximate many quantum kernels in polynomial
time~\cite{Shin2024Dequantizing}; random Fourier feature
approximations provide another classical simulation
pathway~\cite{Sahebi2025Dequantization}; and shadow-based classical
models further compress the quantum-classical
separation~\cite{Jerbi2024Shadows}.  In the other direction, Liu et
al.~\cite{Liu2021rigorous} establish a rigorous end-to-end quantum
advantage by reducing supervised classification to a classically hard
problem (discrete logarithm), but this construction requires fault-tolerant
hardware and applies to a specially constructed dataset --- classical data whose labels
encode discrete-logarithm structure --- not to near-term circuits on natural
data.

The geometric-difference framework of Huang et
al.~\cite{Huang2021Power} provides the standard diagnostic for
quantum-classical kernel separation.  The statistic
$g(K_C, K_Q)$ bounds the worst-case ratio of generalisation errors:
if $g = O(1)$, no target function can benefit from the quantum kernel
by more than a constant factor; if $g \propto \sqrt{N_{\mathrm{train}}}$,
there exist targets where the quantum kernel has an asymptotic advantage.
We apply this diagnostic in Section~\ref{sec:diagnostics} and find, in the
advantage-relevant direction, $g_{CQ} \approx 2.3$--$3.6 \ll
\sqrt{N_{\mathrm{train}}}$ at the primary horizon, consistent with no
asymptotic separation at the tested training sizes.  However, $g > 1$ leaves room for a bounded
finite-sample advantage on specific targets, which is the regime our
empirical results occupy.

Kübler et al.~\cite{Kbler2021Inductive} formalise the conditions under
which a quantum kernel can provide an inductive-bias advantage: the
feature map must realise a low-dimensional RKHS that is aligned with the
target function's structure.  Our target-kernel alignment results
(Section~\ref{sec:diagnostics}, Table~\ref{tab:tka}) are consistent with
this condition: \AeRot{} achieves strictly higher alignment than RBF at
every tested horizon, suggesting that the kernel's RKHS is better matched
to the Lorenz prediction target than the isotropic Euclidean geometry of
RBF.

\section{Experimental Details}
\label{app:details}

\paragraph{Lorenz-63 integration.}
We consider Lorenz system with standard parameters: $\sigma=10$, $\rho=28$, $\beta=8/3$.
For each of the 100 seeds, trajectories are integrated with a fourth-order
Runge--Kutta scheme at $\mathrm{d}t = 0.025$.  The first
$N_{\rm transient} = 1000$ steps are discarded to ensure the trajectory has
settled on the attractor; the subsequent $N_{\rm total} = 2000$ points are
retained.  From each trajectory, $\Ntrain = 80$ training and
$N_{\rm test} = 30$ test windows are extracted.  Two qubit configurations
are evaluated: $n = 5$ (window $N = 32$, $\mathrm{d}t = 0.025$) and
$n = 4$ (window $N = 16$, $\mathrm{d}t_{\rm eff} = 0.05$), chosen so
that the physical prediction times align across configurations.
Training and test sets are drawn from non-overlapping regions of the trajectory, separated by a gap equal to one window length (0.8 tu in physical time) to ensure no raw time-series samples are shared between the last training window and the first test window. The 80 training windows are sampled uniformly from the first 80\% of the available trajectory and the 30 test windows from the remaining 20\%, with the gap applied at the boundary.

\paragraph{Kernel computation.}
For each (quantum, RBF, \matern) kernel method, three per-channel kernels
$K_x$, $K_y$, $K_z$ are built separately and fused via convex
combination $K = \sum_c \beta_c K_c$ with weights selected by inner CV.
Per-channel quantum kernels are computed via exact statevector simulation (no
noise model) using the PennyLane backend.  

\paragraph{CV protocol.}
5-fold inner CV for encoding bandwidth (Rotation angle scaling factor $R$ for \AeRot{}, 
RBF bandwidth $\gamma$ for RBF,
Mat\'{e}rn length scale $\ell$ for Mat\'{e}rn).
Grids: $R \in \{0.177, 0.210, 0.250, 0.297, 0.354\}$; RBF $\gamma = c/N$,
$c \in \{2,\,5,\,8,\,14,\,20\}$; Mat\'{e}rn $\ell = \sqrt{N/(2c)}$ on the same grid.
A second inner CV loop selects the Tikhonov regulariser
$\alpha$ from a logarithmic grid and the three-channel fusion weights
$\beta_x, \beta_y, \beta_z$ ($\beta_i \ge 0$, $\sum_i \beta_i = 1$) by
simplex search on the inner-fold validation loss.
The only structural difference between the classical and quantum sweeps is what produces the kernel matrices (RBF/Mat\'{e}rn formula vs. quantum circuit evaluation). The Hyperparameter selection, beta fusion, and alpha regularization are all selected identically by CV.

\paragraph{Initial guess for $R$}: After per-window $\ell^2$ normalization, $\sum_j \tilde{x}_j^2 = 1$, so the
typical component magnitude is $|\tilde{x}_j| \approx 1/\sqrt{N} = 2^{-n/2}$.
The rotation angle applied by gate $j$ is $\theta_j = \tilde{x}_j / R$.
Setting $\theta_{\rm typical} = 1$\,radian, the maximally nonlinear regime of
$\sin\theta$ (away from the near-linear regime near $0$ and short of
phase-wrapping beyond ${\sim}\pi/2$), gives
$\Rini = 1/\sqrt{N}$. For $n=4$, $\Rini = 0.25$ and for $n=5$, $\Rini \approx 0.177$.

\paragraph{Jacobian computation.}
The $3\times 3$ analytic Lorenz Jacobian is evaluated at each of the $n_{\rm tail}$
raw (unscaled) trajectory points in a window tail using
\texttt{numpy.linalg.eigvals}.  The tail-window stability metric
$\bar{\lambda}_{\rm tail}(n_{\rm tail})$ is the mean of the largest real
eigenvalue over the last $n_{\rm tail}$ steps of each window.

\paragraph{z-scoring conventions.}
Phase-space scatter plot (Fig.~\ref{fig:phase_scatter}, Section~\ref{sec:regime}): 
tail-mean position is computed in
\emph{globally} z-scored coordinates using the empirical mean and
standard deviation of a 200k-step reference trajectory, so the
butterfly background, fixed-point markers, and window markers all
share one frame.
Decile binning by $|x_{\rm tail}|$ (Section~\ref{app:tailfull}): tail-mean $x$ is
computed in \emph{per-seed} z-scored coordinates, statistics of each
seed's $N_{\rm total}=2000$ trajectory.

\section{Extended Horizon and 4-qubit Results}
\label{app:extended}

\begin{figure}[h]
  \centering
   \includegraphics[width=\linewidth]{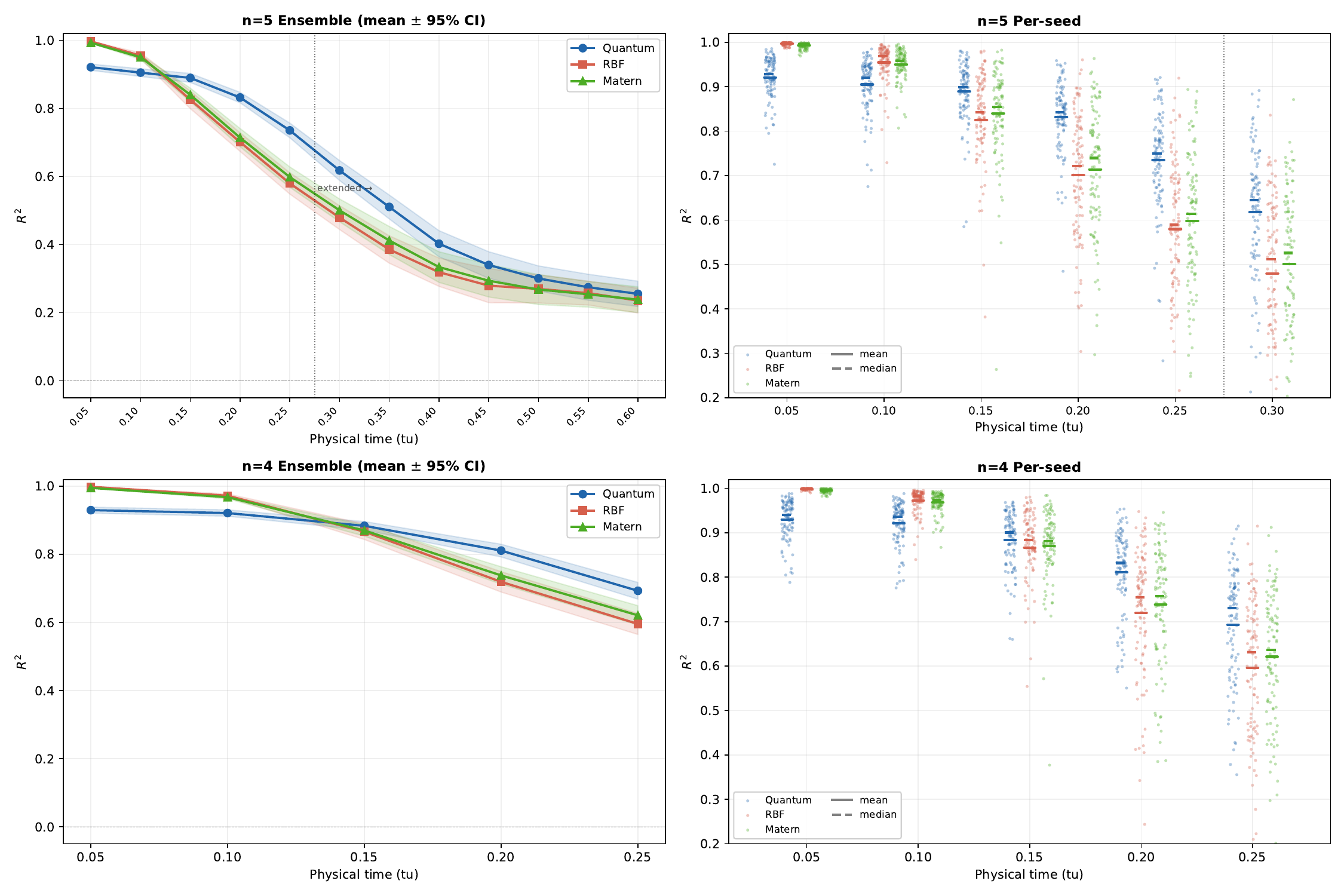}
  \caption{Combined 2x2 figure comparing $n=4$ and $n=5$ configurations side-by-side.
    Top panels: $n=5$ ensemble mean (left) and per-seed scatter (right).
    Bottom panels: $n=4$ ensemble mean (left) and per-seed scatter (right).
    Both configurations show the same qualitative three-phase pattern (no advantage,
    growing advantage, joint collapse), with the crossover between classical and
    quantum-kernel advantage occurring near $0.15$\,tu for both. At 0.20\,tu the configurations
    perform comparably; at 0.25\,tu the $n=5$ advantage emerges against a marginally
    harder prediction task (due to the 0.025\,tu target-time offset).}
  \label{fig:combined_4q_5q}
\end{figure}

Post-hoc regression on pre-computed Gram matrices extends the 5q evaluation
to $0.60$\,tu.
The \AeRot{} advantage is positive through $\approx 0.45$\,tu
($\approx 41\%$ of $\TL$) and narrows to near-zero by $0.50$\,tu as both
kernels degrade together.
Beyond $0.45$\,tu the advantage is not statistically robust.
Figure.~\ref{fig:combined_4q_5q} shows $R^2$ for $n=4$ and $n=5$ with extended horizons for $n=5$.

\section{L2 normalization vs.\ Global Min-Max Scaling}
\label{app:l2norm}

Per-window $\ell^2$ normalization (L2) is compared against global per-trajectory
min-max scaling (rotnorm), which maps each value into $[0, \pi]$ using
training-set extrema.
At $n=4$, $R=0.25$, $n_{\rm ahead}=4$ ($0.20$\,tu), 20 seeds:

\begin{center}
\small
\begin{tabular}{lcc}
  \toprule
  normalization & Mean $R^2$ & Std \\
  \midrule
  Per-window L2 ($R=0.25$) & \textbf{0.862} & 0.067 \\
  Global min-max, best $R$  & 0.801          & 0.085 \\
  \bottomrule
\end{tabular}
\end{center}

Per-position min-max calibrates to training-trajectory extrema.
A window traversing a low-variance orbit segment maps into a tiny angle
subset of $[0,\pi]$, placing all its gates in the near-linear regime and
losing sensitivity to micro-dynamics.
Per-window L2 normalization encodes trajectory \emph{shape}: every window
contributes equal energy to the rotation angles regardless of its position
on the attractor.
For chaotic forecasting, the relative shape of the recent history (momentum,
curvature) is more predictive of the immediate future than absolute
coordinate values.

\section{AE-Only Effective Rank}
\label{app:aerank}

The AE-only kernel $\kappa_{\rm AE}(x,y) = (\tilde{x}\cdot\tilde{y})^2$
(squared cosine similarity) is purely classical.
Its effective rank $\mathcal{R}_{\rm eff} = (\sum_i\lambda_i)^2/\sum_i\lambda_i^2$
measures geometric diversity of training windows in the $\ell^2$-normalized
embedding before the rotation layer adds structured encoding.
Computed at $n=5$, $\Ntrain=80$, 20 seeds:

\begin{table}[!h]
  \centering
  \small
  \caption{AE-only effective rank per channel ($n=5$, $\Ntrain=80$, 20 seeds).}
  \begin{tabular}{lccc}
    \toprule
    Channel & Mean $\mathcal{R}_{\rm eff}$ & Std & Range \\
    \midrule
    $x$ & 4.06 & 0.30 & 3.49--4.60 \\
    $y$ & 5.33 & 0.37 & 4.58--5.95 \\
    $z$ & 5.07 & 0.24 & 4.70--5.73 \\
    \bottomrule
  \end{tabular}
    \label{tab:ae_effrank}
\end{table}

All channels exceed the Lorenz-63 attractor dimension $D \approx 2.06$,
confirming the AE embedding captures the physical dimensionality.

\section{Dominance of $K_y$}
\label{app:Ky}
Both quantum and RBF kernels operate on three Lorenz channels
$(x,y,z)$ independently, producing per-channel $80\times 80$ Gram matrices.
The fused kernel used for prediction is
$K_\mathrm{fused} = \sum_{c} \beta_c K_c$,
where $\beta=(\beta_x,\beta_y,\beta_z)$ is selected by inner CV.
Table~\ref{tab:betas} shows the CV-selected fusion weights at the primary
horizon (0.20\,tu, 100 seeds for both configs).
Because $\beta_y \approx 0.95$ for the quantum kernel and $\beta_y = 1$ for
RBF, the comparison throughout this study is effectively
$K_{y,\mathrm{scr}}$ vs.\ $K_{y,\mathrm{rbf}}$.
The physical rationale is that $\dot{y} = x(\rho-z)-y$ carries multiplicative
cross-channel coupling, so a short $y$-window encodes more joint state
information than a comparable $x$ or $z$ window; while Takens' theorem
guarantees any single generic observable is sufficient given a long enough
window, the $y$-channel is the most efficient choice at the short window
lengths we use.

\begin{table}[!h]
\caption{Mean $\pm$ std of CV-selected fusion weights at 0.20\,tu (100 seeds
each config).  Both kernels reduce to the $y$-channel for all configurations;
the near-zero $\beta_z$  is even more extreme for 5q.}
\centering
\begin{tabular}{llccc}
\toprule
Config & Kernel & $\bar\beta_x$ & $\bar\beta_y$ & $\bar\beta_z$ \\
\midrule
4q & Quantum & $0.095\pm0.164$ & $0.872\pm0.180$ & $0.033\pm0.059$ \\
4q & RBF     & $0.000\pm0.000$ & $1.000\pm0.000$ & $0.000\pm0.000$ \\
\midrule
5q & Quantum & $0.066\pm0.161$ & $0.926\pm0.164$ & $0.008\pm0.034$ \\
5q & RBF     & $0.000\pm0.000$ & $1.000\pm0.000$ & $0.000\pm0.000$ \\
\bottomrule
\end{tabular}

\label{tab:betas}
\end{table}

\section{Hyperparameter Selection Shift with Horizon}
\label{app:hpshift}

\begin{figure}[h]
  \centering
   \includegraphics[width=\linewidth]{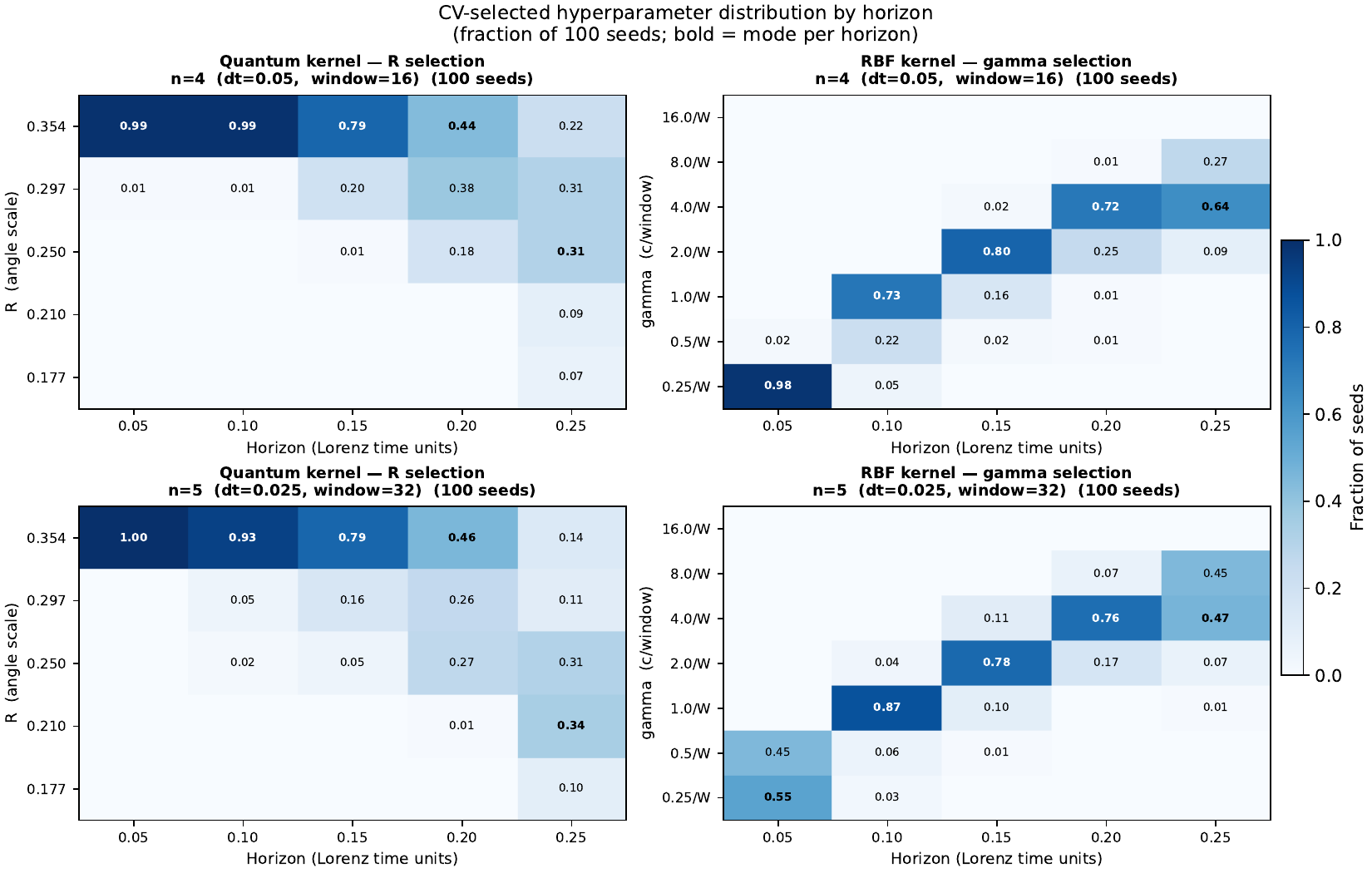}
  \caption{CV-selected hyperparameter distribution by horizon (fraction of
    100 seeds). Left: quantum $R$ (larger = smoother); hot mass shifts
    downward with increasing horizon. Right: RBF $\gamma$ (larger =
    tighter); hot mass shifts upward. Both reflect the same underlying
    shift toward sharper, more local kernels as task difficulty increases.}
  \label{fig:hp-shift}
\end{figure}

\begin{figure}[!h]
  \centering
   \includegraphics[width=\linewidth]{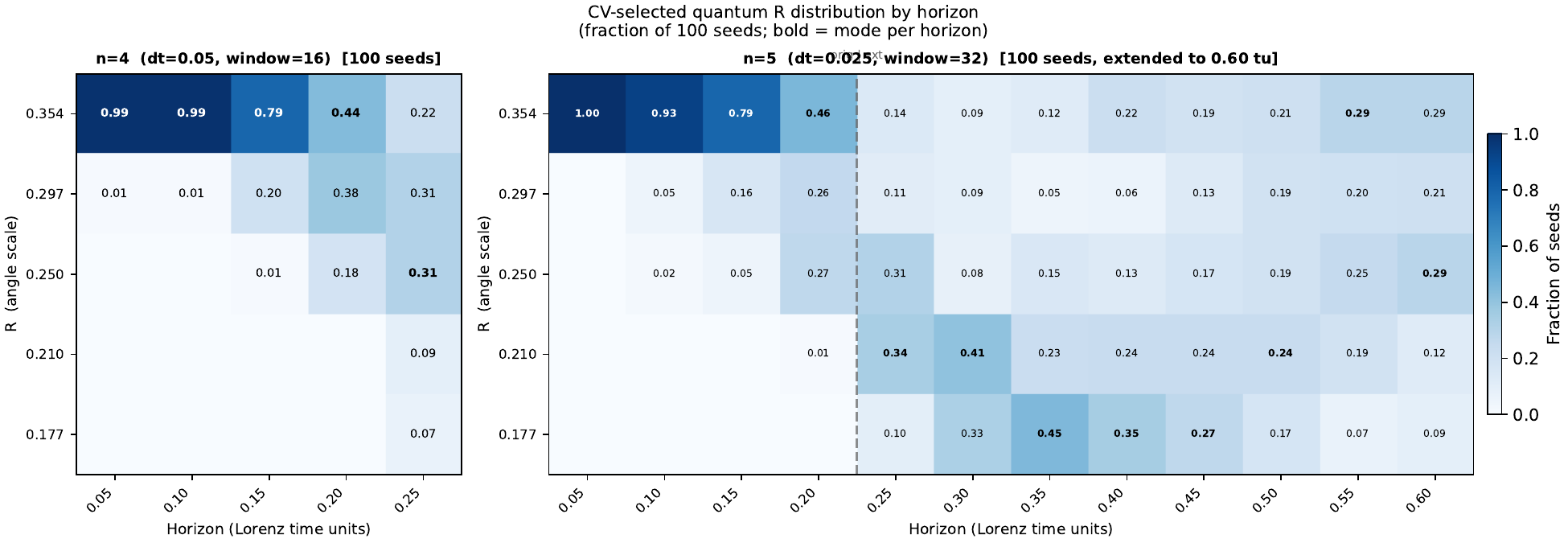}
  \caption{Extended-horizon CV-selected $R$ distribution for 5q
    ($0.30$--$0.60$\,tu). The dominant operating point shifts to
    $R=0.177$ by $0.35$\,tu and disperses by $0.50$\,tu.}
  \label{fig:hp-shift-ext}
\end{figure}

Both the quantum kernel angle scale $R$ and the RBF bandwidth $\gamma$ are
tuned independently per seed and horizon by 5-fold inner CV.
Figures~\ref{fig:hp-shift} and~\ref{fig:hp-shift-ext} visualize the fraction
of seeds selecting each hyperparameter value at each horizon; the RMSE-by-$R$
breakdown is in Table~\ref{tab:rmse}.

\paragraph{Quantum $R$: shifts toward smaller values at longer horizons.}
Larger $R$ produces smaller encoding angles $\tilde{x}_j/R$, making the
kernel smoother and more global; smaller $R$ produces a tighter, more
locally discriminative kernel.
At $0.05$--$0.10$\,tu, $99$--$100\%$ of seeds select $R=0.354$ for both
configurations, consistent with near-linear short-horizon dynamics.
At $0.20$\,tu the operating region spreads and spans $R=0.250$--$0.354$.
At $0.25$\,tu for 5q, seed mass concentrates at $R=0.210$ (34 seeds, mean
RMSE $0.465$) and $R=0.250$ (31 seeds, RMSE $0.478$); seeds still selecting
$R=0.354$ have worse mean RMSE $0.573$, confirming the operating point has
shifted.
The RMSE correlation with CV-selected $R$ at $0.25$\,tu is positive and
significant: $r=+0.397$ ($p<0.001$) for 5q and $r=+0.227$ ($p=0.023$) for 4q.

\paragraph{Extended horizons (5q).}
At $0.30$\,tu the dominant region shifts further: $R=0.177$ (33 seeds, RMSE
$0.549$) and $R=0.210$ (41 seeds, RMSE $0.585$).
By $0.50$\,tu no single $R$ value exceeds 25 seeds, indicating that the
stable operating point disperses beyond the main advantage regime.

\paragraph{RBF $\gamma$: shifts toward larger values at longer horizons.}
Larger $\gamma$ gives a shorter effective length scale, corresponding to
finer local discrimination.
At $0.05$\,tu almost all seeds select the smallest $\gamma = 2/W$; by
$0.20$\,tu the mode has moved to $4/W$ (a 16-fold increase in $\gamma$).
At $0.25$\,tu for 5q, values $4/W$ and $8/W$ share the mass nearly equally
($47\%$ and $45\%$), approaching the upper grid boundary.
By looking at extended $\gamma$ range, we confirm that the upper boundary of the original grid did not pose a limit on the performance at longer horizons, hence the quantum-kernel advantage is not an artifact of insufficient tuning.

\paragraph{Physical interpretation.}
The two trends are physically consistent: large $R$ (smoother quantum
encoding) is analogous to small $\gamma$ (wider RBF kernel), and both
correspond to a more global, less discriminative kernel.
At short horizons the prediction task is locally near-linear and CV favors
the smoothest kernel in each family; as the horizon grows and the predictand
varies more rapidly, CV shifts both kernels toward finer local resolution.
This shift is a direct manifestation of the bias--variance trade-off driven
by task difficulty, and confirms that CV is selecting operating points that
are physically interpretable rather than arbitrary.

\begin{table}[!h]
  \centering
  \small
  \begin{tabular}{l r ccccc}
    \toprule
    Config & Horizon & $R=0.177$ & $R=0.210$ & $R=0.250$ & $R=0.297$ &
      $R=0.354$ \\
    \midrule
    n=4 & 0.05\,tu &  ---        &  ---        &  ---        &  ---        & 0.246\,(99) \\
        & 0.10\,tu &  ---        &  ---        &  ---        &  ---        & 0.262\,(99) \\
        & 0.15\,tu &  ---        &  ---        &  ---        & 0.326\,(20) & 0.321\,(79) \\
        & 0.20\,tu &  ---        &  ---        & 0.390\,(18) & 0.402\,(38) & 0.429\,(44) \\
        & 0.25\,tu & 0.531\,(7)  & 0.519\,(9)  & 0.487\,(31) & 0.537\,(31) & 0.581\,(22) \\
    \midrule
    n=5 & 0.05\,tu &  ---        &  ---        &  ---        &  ---        & 0.262\,(100) \\
        & 0.10\,tu &  ---        &  ---        &  ---        & 0.247\,(5)  & 0.289\,(93)  \\
        & 0.15\,tu &  ---        &  ---        & 0.371\,(5)  & 0.330\,(16) & 0.306\,(79)  \\
        & 0.20\,tu &  ---        &  ---        & 0.408\,(27) & 0.379\,(26) & 0.385\,(46)  \\
        & 0.25\,tu & 0.447\,(10) & 0.465\,(34) & 0.478\,(31) & 0.536\,(11) & 0.573\,(14)  \\
        & 0.30\,tu & 0.549\,(33) & 0.585\,(41) & 0.614\,(8)  & 0.684\,(9)  & 0.655\,(9)   \\
        & 0.35\,tu & 0.643\,(45) & 0.695\,(23) & 0.715\,(15) & 0.710\,(5)  & 0.657\,(12)  \\
        & 0.40\,tu & 0.704\,(35) & 0.753\,(24) & 0.770\,(13) & 0.762\,(6)  & 0.786\,(22)  \\
        & 0.45\,tu & 0.779\,(27) & 0.790\,(24) & 0.743\,(17) & 0.778\,(13) & 0.834\,(19)  \\
        & 0.50\,tu & 0.817\,(17) & 0.779\,(24) & 0.780\,(19) & 0.831\,(19) & 0.841\,(21)  \\
    \bottomrule
  \end{tabular}
  \caption{Mean test RMSE by CV-selected $R$, all horizons, 100 seeds.
    Parentheses show seed count in each $R$ group.
    ``---'' indicates fewer than 3 seeds selected that $R$.
    The CV-dominant operating region ($\geq 25$ seeds) shifts from
    $R=0.354$ at short horizons to $R=0.177$--$0.210$ at long horizons
    for 5q. Extended horizons (below the rule) are post-hoc evaluations
    on pre-computed Gram matrices.}
  \label{tab:rmse}
\end{table}

\section{Structural Diagnostics: Full Tables}
\label{app:diagnostics}

Tables~\ref{tab:oppoint4q} and~\ref{tab:oppoint5q} report four diagnostics
at the CV-optimal operating point across all five prediction horizons for 4q
and 5q respectively.
$\mathrm{Var}_\mathcal{D}$, $\eta_{\rm max}$, and $\varepsilon_U$ are the
kernel entry variance, leading normalized eigenvalue, and expressivity
measure~\cite{FlrezAblan2025similarity};
$F$ is the normalized Frobenius distance.  (The per-horizon
geometric-difference columns and the $g(\Ntrain)$ scaling table of the
submitted version are omitted here: those values were computed in the
reversed direction $g(K_Q \Vert K_C)$; corrected-direction values at the
primary horizon appear in Section~\ref{sec:diagnostics}, and the scaling
analysis awaits recomputation in the corrected direction.)

\begin{table}[!h]
\caption{Operating-point diagnostics (4q, 100 seeds, median across seeds).
    $^\dagger$Primary horizons.}
  \centering
  \small
  \begin{tabular}{r ccc ccc c c}
    \toprule
    & \multicolumn{3}{c}{$\mathrm{Var}_\mathcal{D}[\kappa]$}
    & \multicolumn{3}{c}{$\eta_{\rm max}$}
    & & \\
    \cmidrule(lr){2-4}\cmidrule(lr){5-7}
    Phys.\ time & Q & RBF & ratio & Q & RBF & ratio
      & $F$ & $\varepsilon_U$ \\
    \midrule
    0.05\,tu & 2.688 & 2.176 & 1.24 & 0.137 & 0.101 & 1.36 & 2.180 & 0.200 \\
    0.10\,tu & 2.684 & 2.176 & 1.23 & 0.136 & 0.101 & 1.35 & 0.822 & 0.201 \\
    0.15\,tu & 2.564 & 2.176 & 1.18 & 0.135 & 0.101 & 1.34 & 0.491 & 0.195 \\
    0.20\,tu$^\dagger$ & 2.243 & 2.176 & 1.03 & 0.123 & 0.101 & 1.22 & 0.436 & 0.178 \\
    0.25\,tu$^\dagger$ & 1.845 & 2.176 & 0.85 & 0.115 & 0.101 & 1.14 & 0.445 & 0.160 \\
    \bottomrule
  \end{tabular}
  
  \label{tab:oppoint4q}
\end{table}

\begin{table}[!h]
\caption{Operating-point diagnostics (5q, 100 seeds, median across seeds).
    $^\dagger$Primary horizons.}
  \centering
  \small
  \begin{tabular}{r ccc ccc c c}
    \toprule
    & \multicolumn{3}{c}{$\mathrm{Var}_\mathcal{D}[\kappa]$}
    & \multicolumn{3}{c}{$\eta_{\rm max}$}
    & & \\
    \cmidrule(lr){2-4}\cmidrule(lr){5-7}
    Phys.\ time & Q & RBF & ratio & Q & RBF & ratio
      & $F$ & $\varepsilon_U$ \\
    \midrule
    0.05\,tu  & 1.873 & 2.176 & 0.86 & 0.095 & 0.101 & 0.94 & 2.667 & 0.164 \\
    0.10\,tu  & 1.773 & 2.176 & 0.81 & 0.095 & 0.101 & 0.94 & 1.231 & 0.161 \\
    0.15\,tu  & 1.733 & 2.176 & 0.80 & 0.093 & 0.101 & 0.92 & 0.716 & 0.158 \\
    0.20\,tu$^\dagger$ & 1.511 & 2.176 & 0.69 & 0.086 & 0.101 & 0.85 & 0.462 & 0.146 \\
    0.25\,tu$^\dagger$ & 1.081 & 2.176 & 0.50 & 0.072 & 0.101 & 0.71 & 0.477 & 0.121 \\
    \bottomrule
  \end{tabular}
  
  \label{tab:oppoint5q}
\end{table}

\section{Full TKA table}
Table.~\ref{tab:tka_full} shows the target-kernel alignment table for both $n=4$ and $n=5$, across horizons.
\begin{table}[!h]
  \caption{Target-kernel alignment $A(K_{\rm fused}, YY^\top)$ (100 seeds,
    median). $A_C$ uses the CV-selected fused RBF kernel.
    \AeRot{} achieves higher alignment than RBF at every (config, horizon)
    pair, with $95$--$100\%$ of seeds in favor.}
  \label{tab:tka_full}
  \centering
  \small
  \begin{tabular}{r l r r r r}
    \toprule
    Phys.\ time & Config & $A_Q$ & $A_C$ & $\Delta = A_Q - A_C$ &
      Frac.\ $A_Q > A_C$ \\
    \midrule
    0.05\,tu & 4q & 0.358 & 0.099 & $+0.260$ & 100\% \\
    0.10\,tu & 4q & 0.343 & 0.227 & $+0.112$ & 100\% \\
    0.15\,tu & 4q & 0.323 & 0.244 & $+0.075$ & 100\% \\
    0.20\,tu & 4q & 0.285 & 0.237 & $+0.047$ & 100\% \\
    0.25\,tu & 4q & 0.249 & 0.225 & $+0.025$ &  95\% \\
    \midrule
    0.05\,tu & 5q & 0.345 & 0.106 & $+0.225$ & 100\% \\
    0.10\,tu & 5q & 0.336 & 0.222 & $+0.112$ & 100\% \\
    0.15\,tu & 5q & 0.314 & 0.241 & $+0.076$ &  99\% \\
    0.20\,tu & 5q & 0.289 & 0.233 & $+0.057$ &  99\% \\
    0.25\,tu & 5q & 0.266 & 0.222 & $+0.046$ & 100\% \\
    \bottomrule
  \end{tabular}
\end{table}

\section{Tail-Window Stability: Full Sweep}
\label{app:tailfull}

\begin{figure}[!h]
  \centering
  \includegraphics[width=\textwidth]{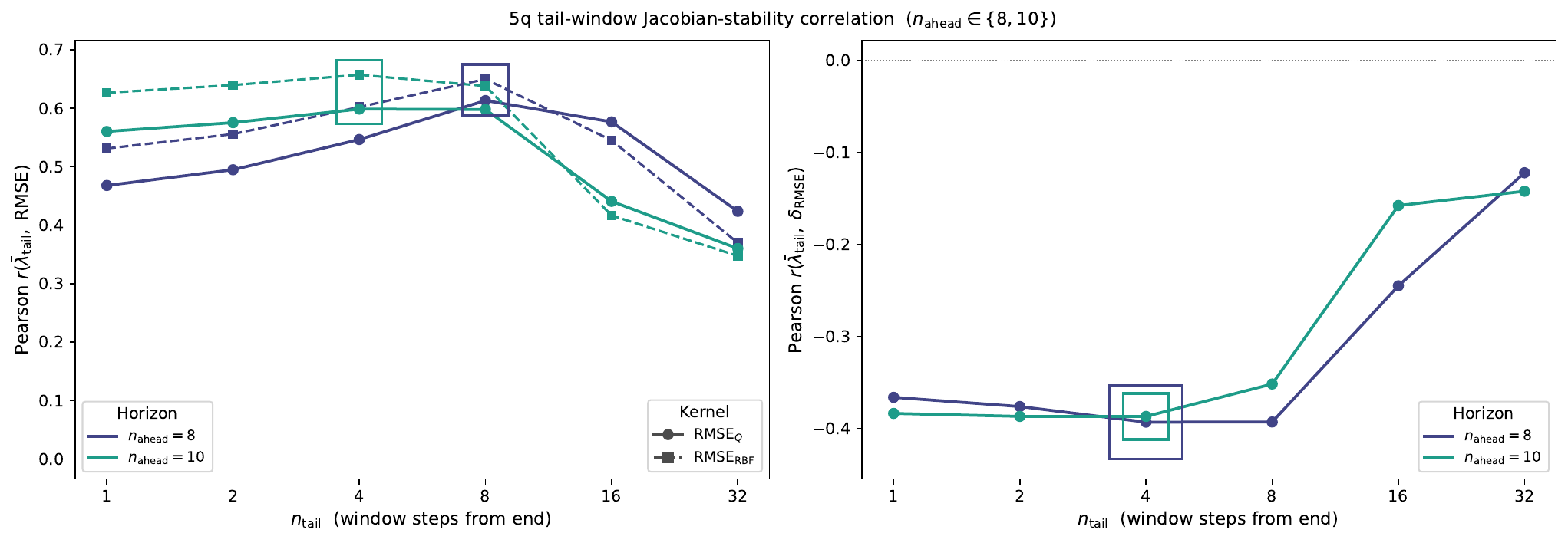}
  \caption{Tail-window stability correlations at $n_{\rm ahead} \in
    \{8, 10\}$ ($0.20$ and $0.25$\,tu).
    \emph{Left}: $r(\bar\lambda_{\rm tail}, \mathrm{RMSE}_{\AeRot})$
    (solid, $\circ$) and $r(\bar\lambda_{\rm tail},\mathrm{RMSE}_{\rm RBF})$
    (dashed, $\square$); same color per horizon.
    \emph{Right}: $r(\bar\lambda_{\rm tail}, \delta_{\rm RMSE})$ (negative =
    \AeRot{} wins).
    colored rectangles mark $n_{\rm tail}^*$.
    RBF curves sit above \AeRot{} curves by a roughly constant positive
    offset at every $n_{\rm tail}$; this offset generates the negative
    $\delta$-correlation in the right panel.}
  \label{fig:tail_n810}
\end{figure}

\begin{figure}[!h]
  \centering
  \includegraphics[width=\textwidth]{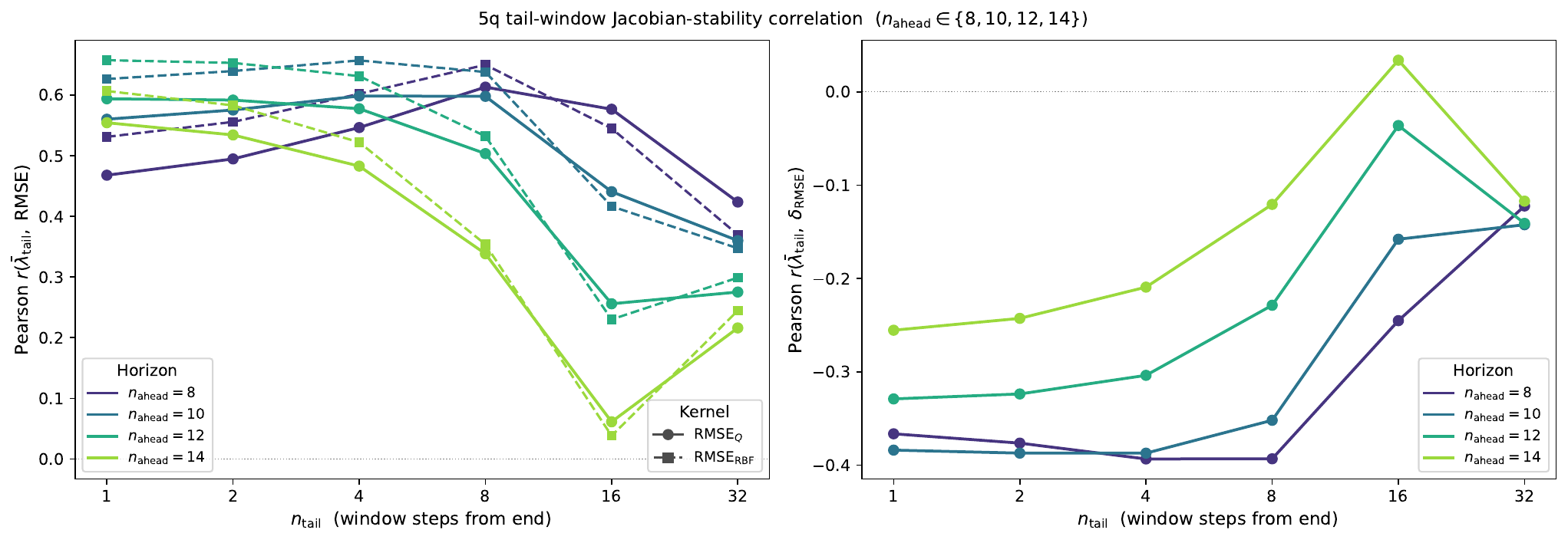}
  \caption{Same panels as Figure~\ref{fig:tail_n810}, extended to all four
    horizons $n_{\rm ahead} \in \{8,10,12,14\}$.
    The per-curve peak (marked $n_{\rm tail}^*$) shifts leftward as the
    horizon grows; both kernels' RMSE correlations and the
    $\delta$-correlation weaken together at longer horizons.}
  \label{fig:tail_all}
\end{figure}

Figures~\ref{fig:tail_n810} and~\ref{fig:tail_all} show the full
$r$-versus-$n_{\rm tail}$ sweep described in Section~\ref{sec:regime}.
Table~\ref{tab:tail_full} reports all correlations numerically.
Table~\ref{tab:tail_win_full} provides Pearson correlation within stable and instable subgroups of all windows, for $\nahead\in\{8,10\}.$

\begin{figure}[!h]
\centering
\includegraphics[width=0.7\textwidth]{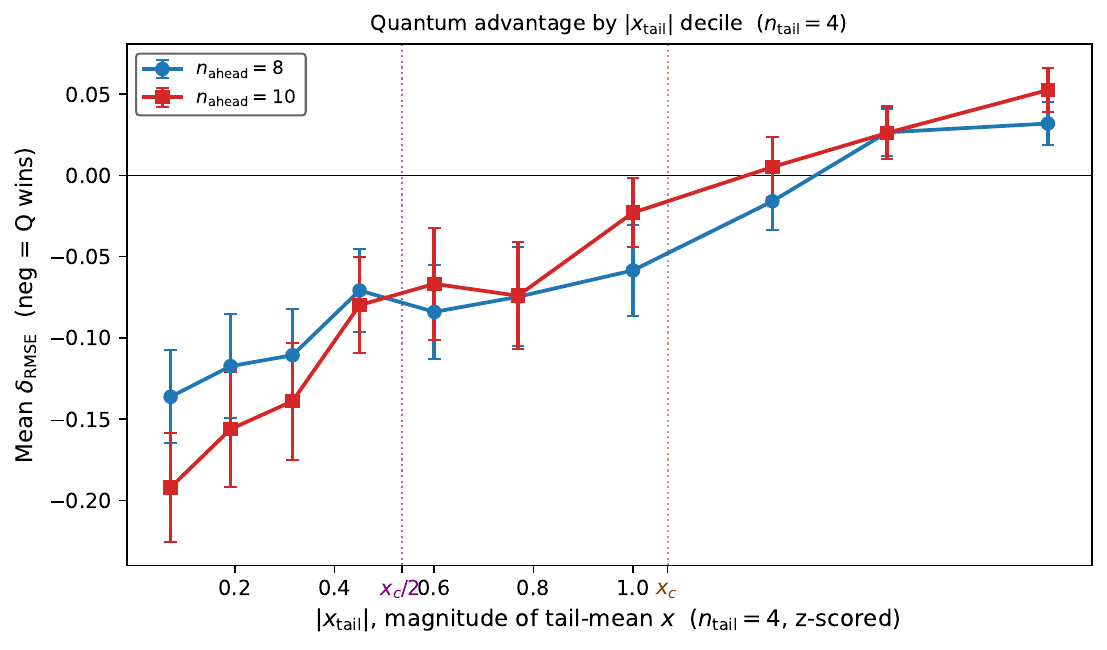}
\caption{Mean $\delta_{\rm RMSE} \pm 2$\,SE per decile of
$|x_{\rm tail}|$, for $n_{\rm ahead} \in \{8, 10\}$.  Monotone:
largest Q advantage at decile~1 (saddle vicinity), fading to zero
near decile~8 and slightly positive at decile~10 (deep lobe).
Verticals at $x_c/2$ and $x_c$ mark the Lorenz attractor's lobe
geometry.}
\label{fig:decile_xtail}
\end{figure}

Figure.~\ref{fig:decile_xtail} provides a complementary view of the location of windows with quantum-kernel advantage.
Binning by $|x_{\rm tail}|$ (mean of $|x|$ over the last $n_{\rm tail}
= 4$ trajectory points, per-seed z-scored) produces a monotone curve:
the largest Q advantage is at decile~1 ($|x_{\rm tail}| \approx 0.07$,
saddle vicinity: $\bar\delta_{\rm RMSE} \approx -0.14$ at
$n_{\rm ahead}=8$, $-0.19$ at $n_{\rm ahead}=10$); the advantage
decays monotonically outward and crosses zero between deciles 8 and 9
($|x_{\rm tail}| \approx 1.3$, just past the lobe centre $x_c =
1.07$); the deepest deciles have $\delta$ slightly positive
(RBF wins by a small margin).

\begin{table}[!h]
\caption{Full sweep of Pearson correlations between
    $\bar\lambda_{\rm tail}$ and per-window RMSE for both kernels and
    their difference $\delta_{\rm RMSE}$ (negative = \AeRot{} wins) at four
    horizons.
    \textbf{Bold} marks the per-column extremum at each horizon.
    $r^*_{\rm RBF} > r^*_{\AeRot}$ at nearly every $(n_{\rm ahead},
    n_{\rm tail})$; the offset is roughly constant in $n_{\rm tail}$
    and drives $r(\bar\lambda_{\rm tail}, \delta_{\rm RMSE}) < 0$.}
  \centering
  \small
  \begin{tabular}{rrrrr}
    \toprule
    $n_{\rm ahead}$ & $n_{\rm tail}$ &
      $r(\bar\lambda_{\rm tail},\,\mathrm{RMSE}_{\AeRot})$ &
      $r(\bar\lambda_{\rm tail},\,\mathrm{RMSE}_{\rm RBF})$ &
      $r(\bar\lambda_{\rm tail},\,\delta_{\rm RMSE})$ \\
    \midrule
     8 &  1 & $+0.468$ & $+0.531$ & $-0.366$ \\
     8 &  2 & $+0.495$ & $+0.556$ & $-0.376$ \\
     8 &  4 & $+0.547$ & $+0.602$ & $\mathbf{-0.393}$ \\
     8 &  8 & $\mathbf{+0.613}$ & $\mathbf{+0.650}$ & $-0.393$ \\
     8 & 16 & $+0.577$ & $+0.545$ & $-0.245$ \\
     8 & 32 & $+0.424$ & $+0.370$ & $-0.122$ \\
    \midrule
    10 &  1 & $+0.560$ & $+0.627$ & $-0.384$ \\
    10 &  2 & $+0.576$ & $+0.640$ & $-0.387$ \\
    10 &  4 & $\mathbf{+0.599}$ & $\mathbf{+0.657}$ & $\mathbf{-0.387}$ \\
    10 &  8 & $+0.598$ & $+0.638$ & $-0.352$ \\
    10 & 16 & $+0.441$ & $+0.417$ & $-0.158$ \\
    10 & 32 & $+0.360$ & $+0.347$ & $-0.142$ \\
    \midrule
    12 &  1 & $\mathbf{+0.594}$ & $\mathbf{+0.658}$ & $\mathbf{-0.329}$ \\
    12 &  2 & $+0.592$ & $+0.653$ & $-0.324$ \\
    12 &  4 & $+0.578$ & $+0.631$ & $-0.304$ \\
    12 &  8 & $+0.504$ & $+0.532$ & $-0.229$ \\
    12 & 16 & $+0.256$ & $+0.230$ & $-0.036$ \\
    12 & 32 & $+0.275$ & $+0.299$ & $-0.141$ \\
    \midrule
    14 &  1 & $\mathbf{+0.555}$ & $\mathbf{+0.607}$ & $\mathbf{-0.255}$ \\
    14 &  2 & $+0.534$ & $+0.583$ & $-0.243$ \\
    14 &  4 & $+0.483$ & $+0.522$ & $-0.209$ \\
    14 &  8 & $+0.338$ & $+0.354$ & $-0.121$ \\
    14 & 16 & $+0.061$ & $+0.038$ & $+0.034$ \\
    14 & 32 & $+0.216$ & $+0.245$ & $-0.117$ \\
    \bottomrule
  \end{tabular}
  
  \label{tab:tail_full}
\end{table}

\begin{table}[!h]
  \caption{Pearson correlation between $\bar\lambda_{\rm tail}$ and
    per-window $\delta_{\rm RMSE} = \mathrm{RMSE}_{\AeRot} - \mathrm{RMSE}_{\rm
    RBF}$ (negative = \AeRot{} wins), split by sign of
    $\bar\lambda_{\rm tail}$ at $n_{\rm ahead} \in \{8, 10\}$.
    The locally stable subgroup shows no significant correlation;
    the unstable subgroup carries essentially all of the overall signal (significant at $p
<10^{-10}$.
    This is not a gradient: the two regimes have qualitatively different
    operative mechanisms.}
  \label{tab:tail_win_full}
  \centering
  \small
  \begin{tabular}{r l r r r r r}
    \toprule
    $n_{\rm ahead}$ & Subgroup & $N$ &
      $r(\bar\lambda_{\rm tail}, \delta_{\rm RMSE})$ & $p$ &
      Mean $\delta_{\rm RMSE}$ & Frac.\ $\delta_{\rm RMSE} < 0$ \\
    \midrule
    8 & Overall                          & 3000 & $-0.393$ & $<10^{-10}$ & $-0.061$ & $54.6\%$ \\
    8 & Stable ($\bar\lambda < 0$)       &  936 & $+0.035$ & $0.29$      & $+0.032$ & $31.2\%$ \\
    8 & Unstable ($\bar\lambda \geq 0$)  & 2064 & $-0.317$ & $<10^{-10}$ & $-0.103$ & $65.2\%$ \\
    \midrule
    10 & Overall                         & 3000 & $-0.387$ & $<10^{-10}$ & $-0.065$ & $54.7\%$ \\
    10 & Stable ($\bar\lambda < 0$)      &  936 & $+0.007$ & $0.83$      & $+0.036$ & $35.0\%$ \\
    10 & Unstable ($\bar\lambda \geq 0$) & 2064 & $-0.322$ & $<10^{-10}$ & $-0.110$ & $63.6\%$ \\
    \bottomrule
  \end{tabular}
  \vspace{-1em}
\end{table}

\begin{figure}[!h]
  \centering
  \includegraphics[width=\textwidth]{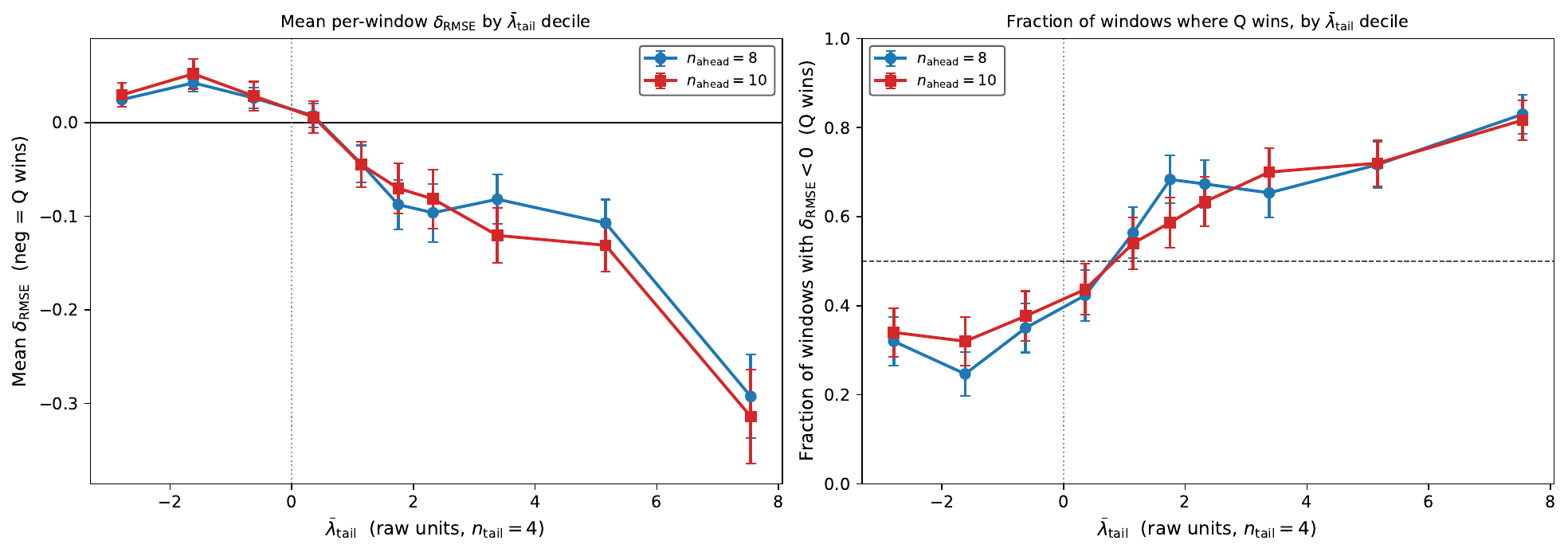}
  \caption{Two-panel decile analysis by $\bar\lambda_{\rm tail}$,
    $n_{\rm ahead} \in \{8, 10\}$.
    \emph{Left}: mean $\delta_{\rm RMSE} \pm 2$\,SE per decile (negative =
    \AeRot{} wins).
    \emph{Right}: per-decile fraction of windows where \AeRot{} wins
    individually ($\delta_{\rm RMSE} < 0$), $\pm 2$\,SE binomial.
    Vertical dotted line at $\bar\lambda_{\rm tail} = 0$ marks the
    locally stable / locally unstable boundary; horizontal dotted line
    at $0.5$ marks parity.
    The \AeRot{} win rate rises from $32\%$ at the most
    stable decile to $83\%$ at the most unstable ($\bar\lambda_{\rm
    tail}$, for both horizons.}
  \label{fig:decile_lambda_full}
  \vspace{-1.2em}
\end{figure}

\section{Echo State Network Comparison}
\label{sec:esn}

\paragraph{Background}
Echo state networks (ESNs)~\cite{Jaeger2004Harnessing} are the standard strong baseline for
Lorenz 63 prediction.  A fixed random recurrent reservoir drives the input trajectory,
and only a linear readout is trained equivalent to KRR with an implicit recurrent
kernel whose feature map is the reservoir state.  Unlike fixed-window KRR, the reservoir
state at time $t$ encodes the \emph{full trajectory history} up to $t$.
Pathak et al.~\cite{Pathak2018Model} showed ESNs reproduce Lorenz 63 attractor statistics,
valid-time predictions, and Lyapunov exponents from data alone.

\paragraph{Configuration and results}
$N_\text{res} = 500$, density $0.1$, fixed reservoir seed $42$.
We consider two CV protocols: a generous \emph{full} protocol that jointly CVs all three hyperparameters, 
and a conventional \emph{$\alpha$-only} protocol that fixes the reservoir and CVs only the ridge penalty.

CV Results (100 seeds) are shown in Table~\ref{tab:esn}.
The ESN is near-perfect at all horizons while the
quantum kernel degrades gracefully and classical fixed-window kernels degrade more steeply.
Bold indicates that ESN has the highest $R^2$ in each row.
Even under the conventional $\alpha$-only protocol the ESN substantially outperforms all
fixed-window methods at every horizon, confirming that the ESN advantage is structural
(full trajectory history) rather than a tuning artifact.

\begin{table}[h]
\caption{CV-corrected mean $R^2 \pm$ std across 100 seeds for all methods.
  ESN (joint): $\rho \in \{0.5,0.9,0.95,0.99\}$, $\sigma \in \{0.1,0.5,1.0,2.0\}$, and $\alpha$ all
  selected jointly by 5-fold inner CV.
  ESN ($\alpha$-only): $\rho=0.9$, $\sigma=1.0$ fixed at standard defaults; only $\alpha$ CV-selected,
  the conventional ESN protocol.
  $n=4$ and $n=5$ cover identical physical horizons (0.05--0.25\,tu) at different
  temporal resolutions (window=16 at $dt=0.05$ vs.\ window=32 at $dt=0.025$).
  Bold marks the highest $R^2$ per row.}
\centering
\small
\resizebox{\textwidth}{!}{%
\begin{tabular}{r r c c c c c}
  \toprule
  n\_ahead & phys\_t & Quantum $R^2$ & RBF $R^2$ & Mat\'{e}rn $R^2$ & ESN (joint) $R^2$ & ESN ($\alpha$-only) $R^2$ \\
  \midrule
  \multicolumn{7}{l}{\textit{n=4 (window=16, $dt_\text{eff}=0.05$)}} \\
  1 & 0.05\,tu & $+0.9299 \pm 0.0449$ & $+0.9985 \pm 0.0017$ & $+0.9952 \pm 0.0041$ & $\mathbf{+1.0000 \pm 0.0000}$ & $+0.9991 \pm 0.0016$ \\
  2 & 0.10\,tu & $+0.9214 \pm 0.0474$ & $+0.9724 \pm 0.0266$ & $+0.9680 \pm 0.0235$ & $\mathbf{+0.9995 \pm 0.0016}$ & $+0.9961 \pm 0.0055$ \\
  3 & 0.15\,tu & $+0.8839 \pm 0.0643$ & $+0.8666 \pm 0.1173$ & $+0.8701 \pm 0.0844$ & $\mathbf{+0.9979 \pm 0.0061}$ & $+0.9903 \pm 0.0115$ \\
  4 & 0.20\,tu & $+0.8111 \pm 0.0953$ & $+0.7193 \pm 0.1529$ & $+0.7381 \pm 0.1302$ & $\mathbf{+0.9800 \pm 0.0265}$ & $+0.9751 \pm 0.0258$ \\
  5 & 0.25\,tu & $+0.6930 \pm 0.1252$ & $+0.5957 \pm 0.1578$ & $+0.6209 \pm 0.1452$ & $\mathbf{+0.9513 \pm 0.0528}$ & $+0.9039 \pm 0.0774$ \\
  \midrule
  \multicolumn{7}{l}{\textit{n=5 (window=32, $dt=0.025$)}} \\
  2  & 0.05\,tu & $+0.9209 \pm 0.0471$ & $+0.9968 \pm 0.0031$ & $+0.9928 \pm 0.0062$ & $\mathbf{+1.0000 \pm 0.0000}$ & $+0.9991 \pm 0.0013$ \\
  4  & 0.10\,tu & $+0.9050 \pm 0.0580$ & $+0.9548 \pm 0.0419$ & $+0.9500 \pm 0.0343$ & $\mathbf{+0.9990 \pm 0.0026}$ & $+0.9966 \pm 0.0036$ \\
  6  & 0.15\,tu & $+0.8894 \pm 0.0670$ & $+0.8253 \pm 0.1327$ & $+0.8397 \pm 0.1011$ & $\mathbf{+0.9952 \pm 0.0110}$ & $+0.9908 \pm 0.0109$ \\
  8  & 0.20\,tu & $+0.8318 \pm 0.0786$ & $+0.7011 \pm 0.1396$ & $+0.7140 \pm 0.1374$ & $\mathbf{+0.9752 \pm 0.0341}$ & $+0.9756 \pm 0.0275$ \\
  10 & 0.25\,tu & $+0.7353 \pm 0.1100$ & $+0.5797 \pm 0.1599$ & $+0.5982 \pm 0.1524$ & $\mathbf{+0.9563 \pm 0.0420}$ & $+0.9314 \pm 0.0566$ \\
  \bottomrule
\end{tabular}%
}

\label{tab:esn}
\end{table}

\paragraph{Scope of the quantum-kernel advantage claim}
The ESN substantially outperforms all fixed-window methods at all horizons
($R^2 > 0.95$ at $0.25$\,tu vs.\ quantum $\approx 0.71$--$0.74$).
This is not a better kernel, it is access to more information: full trajectory history vs fixed window.

The quantum vs.\ classical comparison in this paper is therefore explicitly confined to
the \textbf{fixed-window regime}: methods that see only the last $N$ trajectory points
as input.  Within this regime the quantum-kernel advantage is real, CV-corrected, and
grows with horizon (before collapsing at longer horizons).  Recurrent methods (ESN, LSTM, hybrid
reservoir--physics) sit in a structurally distinct category and are reported here as
context, not as the benchmark the quantum kernel is expected to beat.

\section{Group-Assignment Ablation}
\label{App:group_ablation}
The rotation layer of the \AeRot{} kernel 
involves a design choice: how to assign $2^n$ angles to $n$ qubits.
The assignment scheme changes which time-step
values each qubit encodes and therefore how the kernel measures similarity.

We show that the consecutive-grouped layout, as adopted in all results in the main text, where each qubit encodes a contiguous slice of the window, 
is the best-performing layout among those tested, suggesting that temporal locality of the angle assignment is important for kernel performance.
We demonstrate this by showing that entanglement has opposite effects in the two layouts: 
it improves the block-interleaved layout but degrades the grouped layout, 
yet the blocked-interleaved layout never catches up to the grouped layout at any entangling strength. 

As a controlled example, we consider the \AeRot{} kernel with a brick-wall IsingXX($\theta$) entangling layer and the two arrangement schemes described below.
We sweep the strength of the IsingXX entangling layer from $\theta = 0$ (identity) to $\theta = \pi/2$ (locally equivalent to CNOT). 
Experiments used $n = 4$ qubits, $R = 0.25$, $\Delta t = 0.05$,
$n_\text{ahead} = 4$ ($0.20$ time units), three-channel fusion (x, y, z),
and full simplex search for fusion weights and $\alpha$.
Hyperparameters $\alpha$ and $\beta$ are selected on the test set
(no cross-validation; the experiments are to explore between variant quantum circuit designs).

\subsection{Angle-Assignment Layouts}
\begin{figure}[!h]
  \centering
  \includegraphics[width=0.49\textwidth]{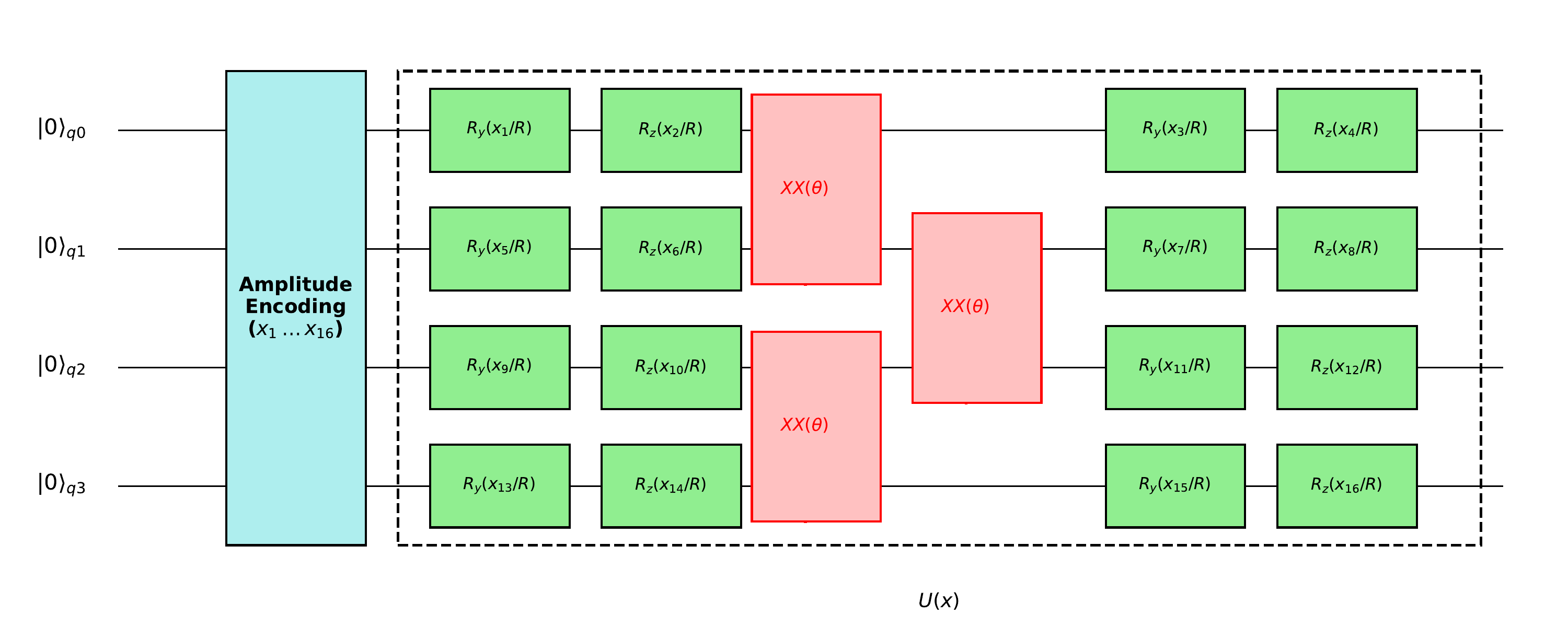}
  \hfill
  \includegraphics[width=0.49\textwidth]{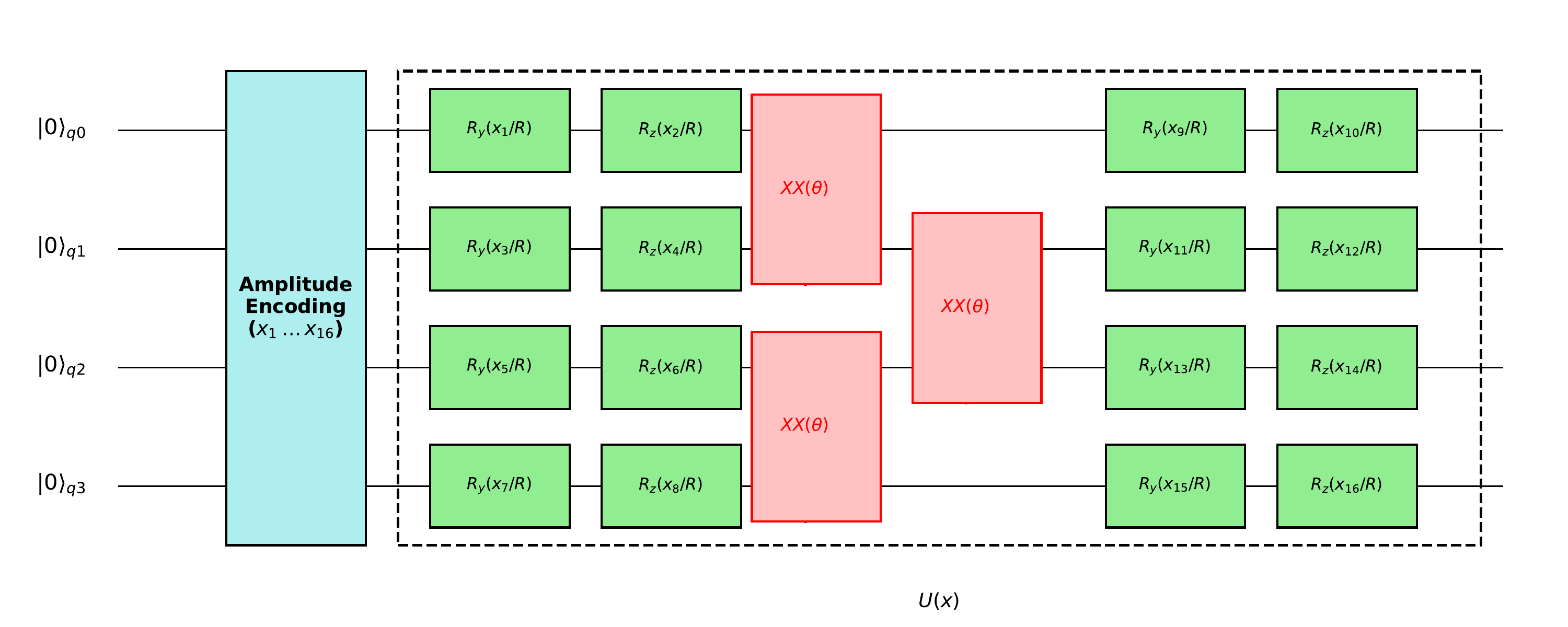}
  \caption{\textit{Left}: Consecutive-grouped layout.  Each qubit encodes a
    contiguous slice: $q_0\!\leftarrow\!\{x_1,\ldots,x_4\}$,
    $q_1\!\leftarrow\!\{x_5,\ldots,x_8\}$, etc.
    \textit{Right}: Block-interleaved layout.  Each qubit encodes two
    non-adjacent pairs: $q_0\!\leftarrow\!\{x_1,x_2,x_9,x_{10}\}$,
    $q_1\!\leftarrow\!\{x_3,x_4,x_{11},x_{12}\}$, etc.
    In both cases the brick-wall IsingXX($\theta$) entangling layer separates
    the two rotation blocks.}
  \label{fig:circuits}
\end{figure}

Figure~\ref{fig:circuits} illustrates the two layouts, Consecutive-grouped (grouped) and Block-interleaved, for $n = 4$ qubits and a window of $N = 16$ steps. 

\paragraph{Consecutive-grouped (default).}
Each qubit owns a contiguous slice of length $\lceil N/n\rceil = 4$:
\[
  q_0 \leftarrow \{1,2,3,4\},\quad q_1 \leftarrow \{5,6,7,8\},\quad
  q_2 \leftarrow \{9,10,11,12\},\quad q_3 \leftarrow \{13,14,15,16\}.
\]

\paragraph{Block-interleaved}
Angles are consumed in qubit-major round-robin order, with $M_\text{rot} = 2$ blocks of $N/M_\text{rot} = 8$ steps each.
\begin{align*}
  &\text{Block 0 (steps 1--8)}:  &q_0 &\leftarrow \{1,2\},\quad q_1 \leftarrow \{3,4\},\quad q_2 \leftarrow \{5,6\},\quad q_3 \leftarrow \{7,8\},\\
  &\text{Block 1 (steps 9--16)}: &q_0 &\leftarrow \{9,10\},\ q_1 \leftarrow \{11,12\},\ q_2 \leftarrow \{13,14\},\ q_3 \leftarrow \{15,16\}.
\end{align*}
Each qubit owns 2 consecutive steps from the first half and 2 from the second half of
the window.  This layout has less temporal locality than the grouped layout, since each qubit's angles are split across two non-adjacent segments of the window.

\subsection{Performance and the effect of entangling}

Table~\ref{tab:layout} shows mean vector-norm $R^2$ across 100 seeds at
$n_\text{ahead} = 4$ ($0.20$ tu), with the IsingXX coupling angle $\theta$
swept from 0 (identity) to $\pi/2$ (locally equivalent to CNOT).
The CNOT baseline uses the standard CNOT gate in place of IsingXX.
Standard deviations range from $\sigma \approx 0.06$ to $\sigma \approx 0.10$, and
95\% confidence intervals on the layout gap (grouped $-$ block-interleaved) are
$\approx \pm 0.012$--$0.014$, confirming that all observed differences are statistically
significant and robust.

\begin{table}[!h]
\caption{Mean vector-norm $R^2$ at $n_\text{ahead}=4$ ($0.20$\,tu) across 100 seeds,
  by IsingXX angle and angle-assignment layout.  B-I = block-interleaved, Grp = grouped.}
\centering
\small
\begin{tabular}{l c c c c c}
  \toprule
  $\theta$ (rad) & Equiv. & B-I mean & B-I std & Grp mean & Grp std \\
  \midrule
  CNOT           & $\pi/2$ & 0.7870 & 0.0942 & 0.8323 & 0.0707 \\
  \midrule
  $0.000$ & identity   & 0.7753 & 0.0912 & \textbf{0.8576} & 0.0611 \\
  $0.050$ &            & 0.7761 & 0.0904 & 0.8569 & 0.0613 \\
  $0.100$ &            & 0.7770 & 0.0898 & 0.8560 & 0.0616 \\
  $0.200$ &            & 0.7792 & 0.0892 & 0.8539 & 0.0621 \\
  $0.300$ &            & 0.7815 & 0.0884 & 0.8516 & 0.0625 \\
  $0.785 \approx \pi/4$ &  & \textbf{0.7882} & 0.0880 & 0.8436 & 0.0637 \\
  $1.047 \approx \pi/3$ &  & 0.7814 & 0.0907 & 0.8398 & 0.0640 \\
  $1.571 \approx \pi/2$ &  & 0.7675 & 0.0998 & 0.8133 & 0.0739 \\
  \bottomrule
\end{tabular}

\label{tab:layout}
\end{table}

\paragraph{Grouped is consistently better.}
The consecutive-grouped layout outperforms block-interleaved regareless of the entangling strength $\theta$, 
with a gap of $\sim +0.04$--$+0.08$ in $R^2$.

\paragraph{Entanglement has opposite effects in each layout.}
\begin{itemize}
  \item \textbf{Grouped}: $R^2$ decreases monotonically with $\theta$,
    from $0.8576$ ($\theta=0$, no entangling) to $0.8133$ ($\theta = \pi/2$),
    a loss of $-0.0443$.
 \item \textbf{Block-interleaved}: $R^2$ increases from $0.7753$ ($\theta=0$) to a peak
    of $0.7882$ ($\theta = \pi/4$), then decreases to $0.7675$ ($\theta = \pi/2$),
    a net gain of $+0.0129$ from identity to maximum entanglement.  
\end{itemize}
In both cases the effect is small relative to the layout gap of $\sim +0.04$--$+0.08$.

\paragraph{Interpretation.}
The un-entangled grouped layout ($\theta=0$) is the best-performing configuration overall.
In the grouped layout each qubit encodes a contiguous temporal segment.
The IsingXX coupling between adjacent qubits mixes information across
time-adjacent segments, which are already correlated by the Lorenz dynamics.
This cross-segment mixing degrades the structured temporal locality that
makes the grouped layout effective.

In the block-interleaved layout each qubit holds two non-adjacent segments
(first- and second-half pairs), so adjacent qubits are less naturally correlated.
While entanglement provides B-I a modest boost by partially recovering temporal correlations
($\theta=0$ to $\pi/4$: +1.29\%), this recovery is fundamentally limited:
The entangling gate create general-purpose correlations, 
but does not have the built-in goal to steer the quantum state 
into a particular subspace that captures the Lorenz dynamics.
The grouped layout's native temporal locality is a more robust and efficient encoding
mechanism—architectural advantage outweighs circuit engineering. 
Even at grouped's minimum (0.8133 at maximum entanglement) exceeds block-interleaved's maximum (0.7882).

We hence conclude that \emph{temporal locality} plays a crucial role in the performance of 
the quantum kernel, and it is more effective to build it in the angle assignment 
than to rely on entanglement to recover it.

\section{Re-uploading Ablation}
\label{app:ablation}
\paragraph{Re-uploading depth and channel-fusion.}
Data re-uploading~\cite{PrezSalinas2020Data} encodes the same input multiple
times through the circuit interleaved with entangling layers, expanding the
accessible Fourier frequency spectrum of the feature map~\cite{Schuld2021effect}
and potentially enriching the kernel.
In our setting, re-uploading with correctly scaled angles
(using $R' = (n_{\rm xx}+1)R$) gives a consistent small gain without the entangling layer; adding entanglement between
re-uploads degrades performance, consistent with the grouped ablation above.

Experiments: $n=4$ qubits, $R=0.25$, $\mathrm{d}t=0.05$, $n_{\rm ahead}=4$
($0.20$\,tu), three-channel fusion, 20 seeds. Hyperparameters $\alpha$ and $\beta$ selected by simplex search optimizing the test $R^2$.  
The goal is to explore circuit options within the quantum setting, not to compare absolute performance against the classical kernels.
IsingXX($\theta$) brick-wall entangling layer swept from $\theta=0$ to
$\theta=\pi/2$. With the correct scaling $R = 0.50 = 2 \times R^\star$, we swept
$\theta \in \{0,\; 0.1,\; 0.2,\; 0.3,\; \pi/8,\; \pi/4,\; \pi/3,\; \pi/2,\; \mathrm{CNOT}\}$
over 20 seeds. Results are in Table.~\ref{tab:reup_4q_20seed}

\begin{table}[!h]
\caption{4-qubit re-uploading kernel, $n_\mathrm{xx}=1$, $R=0.50$: mean $R^2$
vs.\ $\theta$.  19--20-seed means (one seed still running at time of writing).
Bold marks the best configuration.}
\centering
\begin{tabular}{lcc}
\toprule
$\theta$ & Mean $R^2$ & Std \\
\midrule
\AeRot{} (baseline) & +0.862 & 0.066 \\
\midrule
$0$                          & \textbf{+0.875} & 0.059 \\
$0.1$                        & +0.873 & 0.059 \\
$0.2$                        & +0.870 & 0.059 \\
$0.3$                        & +0.868 & 0.059 \\
$\pi/8 \approx 0.393$        & +0.866 & 0.059 \\
$\pi/4 \approx 0.785$        & +0.856 & 0.060 \\
$\pi/3 \approx 1.047$        & +0.850 & 0.065 \\
$\pi/2 \approx 1.571$        & +0.835 & 0.066 \\
CNOT                         & +0.820 & 0.065 \\
\bottomrule
\end{tabular}

\label{tab:reup_4q_20seed}
\end{table}

\paragraph{Re-uploading marginally outperforms scrambling baseline.}
At $\theta = 0$, the 20-seed mean is $+0.875$, a consistent $+0.013$ above the
\AeRot{} baseline of $+0.862$.  This is a small but stable improvement confirmed
across 20 seeds.

\paragraph{Entanglement hurts monotonically.}
Mean $R^2$ decreases strictly as $\theta$ increases from $0$ to $\pi/2$, with
CNOT as the worst performer.  The optimal coupling is $\theta = 0$ (no entanglement).
This mirrors the grouped-assignment \AeRot{} results: interleaved entangling gates
decorrelate the temporally-structured angle pattern.

\paragraph{Re-uploading $\approx$ single encoding at $\theta=0$.}
At $\theta = 0$ and scaled $R$, the circuit is formally
$\mathrm{Rot}(x)^{n_\mathrm{xx}+1}$.
The near-constant performance across re-uploading depths ($n_\mathrm{xx} = 1,2,3$)
is consistent with rotation gates on the same qubit simply composing, making
re-uploading equivalent (up to angle scaling) to a single deeper rotation block.

\noindent Grouped layout: mean $R^2 = 0.875$ at $\theta=0$, degrading
monotonically to $0.820$ at CNOT ($\theta=\pi/2$).\\
Block-interleaved layout: mean $R^2 = 0.761$ at $\theta=0$, improving to
$0.795$ at CNOT.\\
The interleaved layout never matches the grouped layout at any entangling
strength.

\newpage

\end{document}